\documentclass[10pt,journal,compsoc]{IEEEtran}
\usepackage{multirow}
\usepackage{url}
\usepackage{xcolor}
\usepackage{textcomp}
\usepackage{listings}
\usepackage{algorithm}
\usepackage{algorithmic}
\usepackage{subcaption} 
\usepackage{booktabs}  
\usepackage[pdftex]{graphicx}

\ifCLASSOPTIONcompsoc
  \usepackage[nocompress]{cite}
\else
  \usepackage{cite}
\fi
\ifCLASSINFOpdf
\else
\fi
\begin{document}
%
\title{V-Fuzz: Vulnerability-Oriented Evolutionary Fuzzing}
%
%
%
%

\author{Yuwei~Li,
        Shouling~Ji,
        Chenyang~Lv,
        Yuan~Chen,
        Jianhai~Chen,
        Qinchen~Gu,
        and~Chunming~Wu
\IEEEcompsocitemizethanks{
\IEEEcompsocthanksitem Y. Li, S. Ji, C. Lv, Y. Chen, J. Chen and C. Wu are with the College of Computer Science and Technology, Zhejiang University, China.\protect\\
E-mail: \{liyuwei,sji,puppet,chenyuan,chenjh919,wuchunming\}@zju.edu.cn
\IEEEcompsocthanksitem Q. Gu is with the Department
of Electrical and Computer Engineering, Georgia Institute of Technology, Atlanta,
GA, 30332. \protect\\
E-mail: qgu7@gatech.edu
}
}

\IEEEtitleabstractindextext{%
\begin{abstract}
Fuzzing is a technique of finding bugs by executing a software recurrently with a large number of abnormal inputs.\\
Most of the existing fuzzers consider all parts of a software equally,
and pay too much attention on how to improve the code coverage.
It is inefficient as the vulnerable code only takes a tiny fraction of the entire code.
In this paper, 
we design and implement a vulnerability-oriented evolutionary fuzzing prototype named V-Fuzz, 
which aims to find bugs efficiently and quickly in a limited time.
V-Fuzz consists of two main components:  
a \emph{neural network-based vulnerability prediction model}
and a \emph{vulnerability-oriented evolutionary fuzzer}. 
Given a binary program to V-Fuzz, 
the vulnerability prediction model will give a prior estimation on which parts of the software are more likely to be vulnerable.
Then,
the fuzzer leverages an evolutionary algorithm to generate inputs which tend to arrive at the vulnerable locations,  
guided by the vulnerability prediction result. 
Experimental results demonstrate that V-Fuzz can find bugs more efficiently than state-of-the-art fuzzers.
Moreover,
V-Fuzz has discovered 10 CVEs, 
and 3 of them are newly discovered.\\
We reported the new CVEs, 
and they have been confirmed and fixed.
\end{abstract}

}

\maketitle

\IEEEdisplaynontitleabstractindextext

%
\IEEEpeerreviewmaketitle

\IEEEraisesectionheading{\section{Introduction}\label{sec:introduction}}

%
%
%
%
Fuzzing is an automated vulnerability discovery technique by feeding manipulated random or abnormal inputs to a software program
\cite{Sutton2007Fuzzing}.
With the rapid improvement of computer performance, 
fuzzing has been widely used by software vendors such as Google 
\cite{Google} 
and Microsoft 
\cite{MS}
for detecting bugs in softwares.
However, 
the efficiency of fuzzing is still badly in need of being improved, 
especially for detecting bugs for large/complex softwares.

It is time-consuming for fuzzers to discover bugs as they usually work blindly.
The blindness of fuzzers is mainly reflected in the following aspects.
First, 
fuzzers generate inputs blindly. 
Usually, 
the input generation strategies are based on simple evolutionary algorithms,
or even merely mutate the seed inputs randomly without any feedback information,
especially for blackbox fuzzers.
Second, 
fuzzers are blind with the fuzzed software especially for blackbox and greybox fuzzers,
i.e., 
they do not know much information about the fuzzed program.
Blackbox fuzzers 
\cite{Takanen2008Fuzzing}
\cite{Peach} 
treat a program as a black box and are unaware of its internal structure
\cite{Fuzz_wiki}.
They simply generate inputs at random.
Thus, 
discovering a bug is just like looking for a needle in a haystack,
which somewhat relies on luckiness.
Greybox fuzzers are also unaware of the program's source codes.
They usually do some reverse analyses of binary programs.
However,
these analyses are just simple binary instrumentations,
which are mainly used to measure the code coverage.

Although whitebox fuzzers
\cite{Godefroid2008Automated}
\cite{Ganesh2009Taint} 
can perform analysis on the source code of a program to increase code coverage or to reach certain critical program locations,
there are still many problems for whitebox fuzzers to be applied in fuzzing real programs.
Additionally, 
whitebox fuzzers still have blindness.
They usually leverage symbolic execution or similar techniques to generate inputs that try to go through as many paths as possible.
Nevertheless,
symbolic execution-based input generation is largely ineffective against large programs
\cite{Wang2015Experience}.
For instance,
Driller is a guided whitebox fuzzer that combines AFL 
\cite{AFL} 
with concolic execution  
\cite{Driller}.
It was benchmarked with 126 DARPA CGC 
\cite{DARPACGC} binaries,
while its concolic engine can only generate valid inputs for 13 out of 41 binaries. 
Therefore, 
it is not practical to apply symbolic execution to explore meaningful path especially for the large software.
Moreover, 
as it is hard to generate enough valid inputs,
whitebox fuzzers' strategy which blindly pursues high code coverage
is not wise and may waste a lot of time and energy.

As indicated by Sanjay et al.   
\cite{Rawat2017VUzzer},
the crux of a fuzzer is its ability to generate bug triggering inputs.
However,
most of the state-of-the-art fuzzers 
e.g.,
\emph{coverage-based} fuzzers mainly focus on how to improve the code coverage.
Intuitively, 
fuzzers with a higher code coverage can potentially find more bugs.
Nevertheless,
it is not appropriate to treat all codes of the program as equal.
The reasons are as follows.
First, 
it is difficult to achieve a full code coverage for real programs.
As the size of real programs may be very large,
the generation of inputs to satisfy some sanity checks is still extremely hard
\cite{T-Fuzz}.
Second, 
it needs a mass of time and computing resources to achieve high code coverage.
Third, 
the vulnerable code usually takes a tiny fraction of the entire code.
For example,
Shin et al.
\cite{Shin2013can} found that only 3\% of the source code files in Mozilla Firefox have vulnerabilities.
Although higher code coverage can enable a fuzzer to find more bugs, 
most covered codes may not be vulnerable.
Thus,
it is expected to seek a balance between the fuzzing efficiency and the time cost and computing resources.
Therefore, 
fuzzers should prioritize the codes which have higher probability of being vulnerable.

Therefore, 
we develop a vulnerability-oriented fuzzing framework V-Fuzz for quickly finding more bugs on binary programs in a limited time. 
In particular,
we mainly focus on detecting vulnerabilities of binary programs since for most of the cases,
we cannot get access to the source codes of softwares,
especially for the Commercial Off-The-Shelf (COTS) softwares.
Nevertheless,
our approach is also suitable for the softwares which are open source by compiling them into binaries.
V-Fuzz consists of two components: 
a \emph{vulnerability prediction model} 
and a 
\emph{vulnerability-oriented evolutionary fuzzer}.
The prediction model is built based on a neural network.
Given a binary software, 
the prediction model will perform an analysis on which components of the software are more likely to have vulnerabilities. 
Then,
the evolutionary fuzzer will leverage the prediction information to generate high-quality inputs which tend to arrive at these potentially vulnerable components.
Based on the above strategies, 
V-Fuzz reduces the waste of computing resources and time,
and significantly improves the efficiency of fuzzing.
In addition, 
V-Fuzz still gives relatively small attention to other parts of the binary program,
which have a lower probability of being vulnerable.
Therefore, it can mitigate the false negative of the vulnerability prediction model.
In summary, our contributions are the followings:


1) 
We analyze the limitations of existing coverage-based fuzzers.
They usually simply treat all the components of a program as equal,
in hope of achieving a high code coverage,
which is inefficient.

2)
To improve fuzzing efficiency,
we propose V-Fuzz,
a vulnerability-oriented evolutionary fuzzing prototype.
V-Fuzz significantly improves the fuzzing performance leveraging two main strategies:
deep learning based vulnerable component predication and predication information guided evolutionary fuzzing.

3)
To examine the performance of V-Fuzz,
we conduct extensive evaluations leveraging 10 popular Linux applications and three programs of the popular fuzzing benchmark LAVA-M.
The results demonstrate that V-Fuzz is efficient in discovering bugs for binary programs.
For instance,
compared with three state-of-the-art fuzzers 
(VUzzer, AFL and AFLfast),
V-Fuzz finds the most unique crashes in 24 hours.
Moreover,
we discovered 10 CVEs by V-Fuzz, where 3 of them are newly discovered.
We reported the new CVEs, 
and they have been confirmed and fixed.

\section{Background}

\subsection{Motivation of V-Fuzz}

\begin{figure}[!t]
\lstset{frame=single,
	basicstyle=\scriptsize,
	language=C,
	breaklines=true,
	numbers=right,
	numberstyle=\tiny,
	keywordstyle=\color{blue!90},
    commentstyle= \color{red!90}, 
	rulesepcolor= \color{ red!20!green!20!blue!20} ,
}
\begin{lstlisting}[language=C]
#define SIZE 1000
int main(int argc, char **argv){
    unsigned char source[SIZE];
    unsigned char dest[SIZE];
    char *input=ReadData(argv[1]);
    int i;
    i=argv[2];
    /*magic byte check */
    if(input[0]!='*')
        return ERROR;
    if(input[1]==0xAB && input[2]==0xCD){
        printf("Pass 1st check!\n");
        /*some nested condition*/
        if(strncmp(&input[6],"abcde",5)==0){
            printf("Pass 2nd check!\n");
            /* some common codes without vulnerabilities*/
            ...
        }
        else{
            printf("Not pass the 2nd check!");
        }
    }
    else{
        printf("Not pass the 1st check!");     
    }
    /*A buffer overflow vulnerability*/
    func_v(input, source, dest);
    return 0;
}
\end{lstlisting}
\caption{A motivation example.}
\label{motivation}
\end{figure}

Fuzzers can be categorized into several types from different perspectives.
Based on different input generation methods, 
fuzzers can be categorized as generation-based and mutation-based.
A generation-based fuzzer 
\cite{PROTOS}
\cite{SPIKE}
\cite{Peach}
generates inputs according to the input model designed by users.
A mutation-based fuzzer 
\cite{AFL}
\cite{honggfuzz}
\cite{zzuf}
generates inputs based on mutating a corpus of seed files during fuzzing.
In this paper, 
we focus on mutation-based fuzzing.

Based on the exploration strategy, 
fuzzers can be classified as directed and coverage-based.
A directed fuzzer
\cite{Godefroid2005DART} 
\cite{Ganesh2009Taint}
\cite{Haller2013Dowsing} 
\cite{Neugschwandtner2015The}
\cite{AFLGo}
generates inputs that aim to arrive at some specific regions.
A coverage-based fuzzer
\cite{AFL}
\cite{honggfuzz}
\cite{libFuzzer}
\cite{CollAFL}
aims to generate inputs that can traverse as many paths as possible.

Figure \ref{motivation}
shows a highly simplified scenario to describe our motivation.
In this example, 
there is a function 
\texttt{main} 
and it calls a function 
\texttt{func\_v} in line 27. 
There are some magic bytes and error checks in the \texttt{main} function from line 9 to line 25, 
some of which are even nested condition statements. 
It is known that generating inputs which can arrive at the check codes is not easy. 
However, 
the codes for these checks do not have any vulnerability. 
While, 
in line 27, 
the function \texttt{func\_v} has a buffer overflow vulnerability.

For the example in Figure \ref{motivation}, 
most of the state-of-the-art coverage-based fuzzers do not work well.
For instance,
AFL 
\cite{AFL}
is a security-oriented fuzzer that employs an evolutionary algorithm to generate inputs that can trigger new internal states in the targeted binary. 
More specifically, 
if an input discovers a new path, 
this input will be selected as a seed input.
Then,
AFL generates new inputs by mutating the seed inputs.
AFL simply regards the input which discovers a new path as a good input. 
Once a path is difficult to explore, 
AFL gets "stuck" and chooses another path which is easier to arrive.
In this example, 
AFL will spend a lot of time trying to generate inputs that can bypass these checks.
However, 
AFL will easily get "stuck".
Finally,
AFL may find new paths or crashes when it arrives at the path at function \texttt{func\_v}.
Nevertheless, 
the previous effort is barely useful in this case.
Another state-of-the-art fuzzer is VUzzer 
\cite{Rawat2017VUzzer},
which is an application-aware evolutionary fuzzer.
VUzzer is designed to pay more attention to generating inputs which can arrive at deeper paths.
For this example,
VUzzer will give more weights to fuzz the codes nested in the condition statements.
However, 
the codes in deeper paths do not mean that they have a higher probability to be vulnerable.
Therefore, 
the fuzzing strategy of VUzzer does not work well either.

For this example,
the most important parts of the program that fuzzers should concern first is the function \texttt{func\_v}.
If there exists a static analysis tool or vulnerability detection model which can give a warning information that \texttt{func\_v} may have vulnerabilities,
a fuzzer can give more weights to generating inputs which tend to arrive at \texttt{func\_v}.
This will make the fuzzer more efficient in discovering bugs.

Based on the above discussion, 
we propose V-Fuzz, 
which aims to find bugs quickly in a limited time. 
V-Fuzz leverages an evolutionary algorithm to generate inputs which tend to arrive at vulnerable codes of the program, 
assisted by vulnerability prediction information.
It needs to be emphasized that V-Fuzz is neither coverage-based nor directed.
V-Fuzz is different from most coverage-based fuzzers which regard all codes as equal,
since it pays more attention to the codes which have higher probability to be vulnerable.
In addition, 
unlike directed fuzzers such as AFLGo \cite{AFLGo}, 
which generates inputs with the objective of reaching a given set of target program locations,
V-Fuzz gives relatively small weights to other codes which are unlikely to be vulnerable. 
This is because the vulnerability prediction model may not be so accurate, 
and the components that are predicted to be safe may still be vulnerable.
Therefore,
V-Fuzz leverages the advantages of vulnerability prediction and evolutionary fuzzing,
and meanwhile reduces the disadvantages of them.

\subsection{Binary Vulnerability Prediction}

In order to implement a vulnerability-oriented fuzzer,
a vulnerability prediction module is expected,
which performs a pre-analysis of which components are more likely to be vulnerable.
There are two main approaches for vulnerability prediction.
One is using traditional static vulnerability detection methods.
The other is leveraging machine learning or deep learning techniques. 
Among the two approaches, 
we choose to leverage deep learning to build the vulnerability prediction model due to the following reasons.

First, 
most of traditional static analysis methods use pattern-based approaches to detect vulnerabilities.
The patterns are manually defined by security experts,
which is difficult,  
tedious and time-consuming.
For example, 
there are many open source vulnerability detection tools such as 
ITS4  
\cite{Viega2000ITS4},
and commercial tools such as
Checkmarx  
\cite{checkmarx}.
These tools often have high false positive rates or false negative rates
\cite{Li2018VulDeePecker}.
In addition, 
these tools are used to detect vulnerabilities within source codes.
As V-Fuzz focuses on fuzzing binary programs,
these tools are not suitable for V-Fuzz.

Second, 
deep learning has been applied to the fields of defect prediction  
\cite{Yang2015Deep}
\cite{Wang2016Automatically},
program analysis  
\cite{White2015Toward} 
\cite{Shin2015Recognizing}
\cite{Xu2017Neural}
and vulnerability detection 
\cite{Li2018VulDeePecker}
successfully.
In these applications,
deep learning has several advantages when compared with pattern-based methods:
(1) Deep learning methods do not need experts to define the features or patterns,
which can reduce lots of overhead.
(2) Even for the same type of vulnerabilities, 
it is hard to define an accurate pattern that can describe all forms of it.
(3) Pattern-based methods usually can only detect one specific type vulnerability,
while deep learning methods have been proven to be able to detect several types of vulnerabilities simultaneously
\cite{Li2018VulDeePecker}.
Therefore, 
we choose to leverage deep learning methods to build our vulnerability prediction model.

Moreover, 
it is worth to note that there has been no such approach that leverages deep learning to detect or predict vulnerabilities for binary programs to the best of our knowledge.
In order to build our prediction model, 
there are several questions that need to be considered.

\textbf{How to represent a binary program?}
It is inappropriate to analyze binary codes directly,
as the binary itself does not have sufficient syntactic or semantic information. 
Therefore, 
the binary needs to be transformed into some intermediate form such as assembly language,
which has enough meaningful characteristic information. 
Following this idea,
we choose to analyze the \emph{Control Flow Graph (CFG)} of a binary program, 
as the \emph{CFG} contains rich semantic and structural information.
In addition,
in order to conveniently train a deep learning model,
the binary program needs to be represented as numerical vectors.
Therefore, 
we leverage the \emph{Attributed Control Flow Graph (ACFG)} 
\cite{Feng2016Scalable} 
to describe and represent a binary program with numerical vectors.

\textbf{Which granularity is suitable for analysis?}
The granularity is also an important factor that needs to be considered. 
The possible granularity of a binary program can be file, 
function, 
basic block, 
etc. 
In fact,
a proper granularity cannot be too coarse or too fine, 
as the too coarse granularity will decrease the precision of the model, 
while the too fine granularity will make it hard to collect sufficient labeled data for training a meaningful model. 
Therefore, 
we seek a tradeoff  and choose function as the granularity to analyze.

\textbf{Which neural network model is appropriate for the vulnerability prediction problem?}
One important advantage of neural networks is that it can learn features automatically.
Since we leverage \emph{ACFG} to represent a binary program,
we leverage graph embedding network 
\cite{Dai2016Discriminative} to build our model
as it has been successfully applied to extract valid features of structural data.
Moreover,
Xu et al.  
\cite{Xu2017Neural} applied this approach to detect similar cross-platform binary codes.
Therefore, 
it could be feasible to leverage this model to predict the vulnerable components of a program.
As the aim of our model is not detecting similar codes,
it is necessary to change the graph embedding network to make it suitable for our problem.
The detailed description will be presented later.

\section{V-Fuzz: System Overview}
\begin{figure}[!t]
\centering
\includegraphics[width=3.5in]{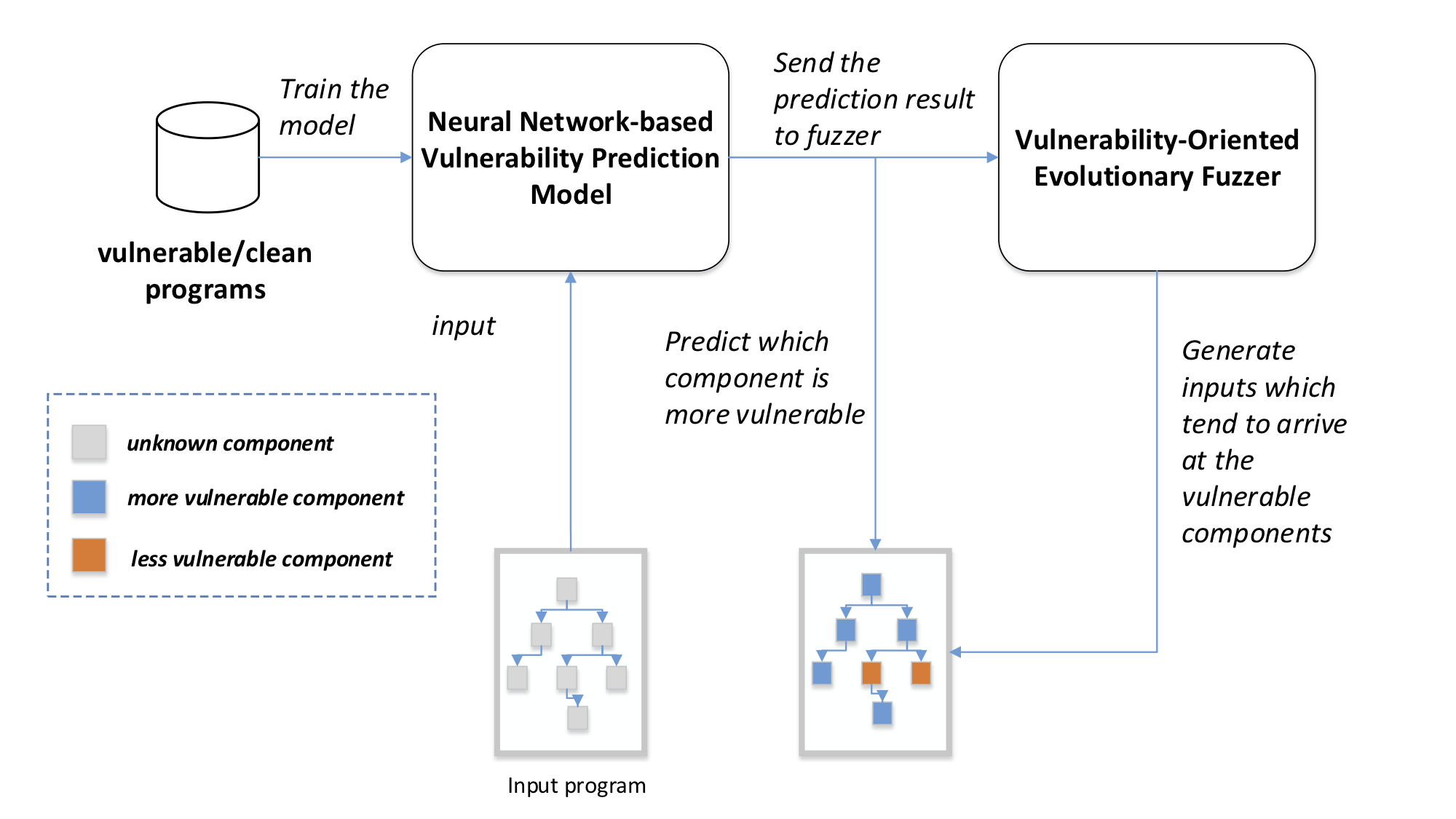}
\caption{The architecture of V-Fuzz.}
\DeclareGraphicsExtensions.
\label{archit-v}
\end{figure}

In this section, 
we introduce the main components and workflow of V-Fuzz.
Figure \ref{archit-v} shows the architecture of V-Fuzz,
which consists of two modules: 
a \emph{Neural Network-based Vulnerability Prediction Model} 
and a \emph{Vulnerability-Oriented Evolutionary Fuzzer}.

\textbf{Neural Network-based Vulnerability Prediction Model.}
For a binary program, 
the vulnerability prediction model will give a prediction on which components are more likely to be vulnerable.
More specifically,
the component is a function of a binary program,
and the prediction is the probability of a function being vulnerable.
We leverage deep learning to build this model,
and the core structure of the model is a graph embedding network.
The detailed structure of the model is shown in Section 4.
In order to enable the model to predict vulnerabilities,
we train the model with a number of labeled data (the label is "vulnerable" or "secure").
In addition,
the model is able to predict several types of vulnerabilities when it is trained with sufficient data related to these vulnerabilities.
Then,
the prediction result will be sent to the fuzzer to assist it in finding bugs.

\textbf{Vulnerability-Oriented Evolutionary Fuzzer.}
Based on the previous vulnerability prediction result,
the fuzzer will assign more weight to the functions that have higher vulnerable probabilities.
For convenience, 
we use the "vulnerable probability" to represent the "the probability of being vulnerable" in this paper.
The process is as follows:
for each function of the binary program with a vulnerable probability,
V-Fuzz will give each basic block in the function a \emph{Static Vulnerable Score (SVS)}, 
which represents the importance of the basic block.
The detailed scoring method is described in Section 5.
Then, 
V-Fuzz starts to fuzz the program with some initial inputs provided by users.
It leverages an evolutionary algorithm to generate proper inputs.
For each executed input,
V-Fuzz gives a \emph{fitness score} for it,
which is the sum of the \emph{SVS}
of all the basic blocks that are on its execution path.
Then, 
the inputs that have higher \emph{fitness scores} or cause crashes will be selected as new seed inputs.
Finally,
new inputs will be continuously generated by mutating the seed inputs.
In this way, 
V-Fuzz tends to generate inputs that are more likely to arrive at the vulnerable regions.

\section{Vulnerability Prediction}

\subsection{Problem Formalization}

In this subsection, 
we formalize the vulnerability prediction problem. 
We denote the vulnerability prediction model as $M$. 
Given a binary program $\hat{p}$, 
suppose it has $\tau$ functions 
$F={\{f_{1}, f_{2}, ... , f_{\tau}\}}$.
For any function $f_{i} \in F$,
it is an input of
$M$, 
and the corresponding output $PV_{f_{i}}$ 
denotes the vulnerable probability of $f_{i}$,
i.e.,
\begin{equation}
PV_{f_{i}}=M(f_{i}).
\label{1}
\end{equation}

The function that has a high vulnerable probability should be paid more attention when fuzzing the program. 
In order to build such an $M$, 
there are three aspects to be considered: 
\emph{the representation of input data}, i.e.,
the approach of data preprocess,
\emph{the model structure} 
and
\emph{how to train and use the model}.
We will introduce the details of the three aspects below.

\subsection{Data Preprocessing}

As discussed in Section 2,
to build and train $M$,
we should seek a method to transform binary program functions into numerical vectors.
Moreover, 
the vectors should be able to carry enough information for future training.
Towards this,
we choose to leverage the 
\emph {Attributed Control Flow Graph (ACFG)} 
\cite{Feng2016Scalable} 
to represent the binary function.

{\bfseries\emph{ACFG}} is a directed graph 
$g=<V,E,  \phi>$,
where $V$ is the set of vertices, 
$E$ is the set of edges,
and $\phi$ is a mapping function.
In \emph{ACFG}, 
a vertex represents a basic block,
an edge represents the connection between two basic blocks,
and
$\phi:V\rightarrow\sum$ 
maps a basic block in $g$ to a set of attributes 
$\sum$.

As we know,
it is common to use \emph{Control Flow Graph (CFG)} to find bugs
\cite{pewny2015cross}
\cite{Eschweiler2016discovRE}.
However,
\emph{CFG} is not a numerical vector,
which means it cannot be used to train a deep learning model directly.
Fortunately, 
\emph{ACFG} is another form of \emph{CFG} by describing \emph{CFG} with a number of basic-block level attributes.
In \emph{ACFG},
each basic block is represented by a numerical vector,
where each dimension of the vector denotes the value of a specific attribute.
In this way,
the whole binary function can be represented as a set of vectors.
Therefore,
\emph{ACFG} is suitable for our requirements to represent a binary function.

\begin{figure}[!t]
\centering
\includegraphics[width=3.5in]{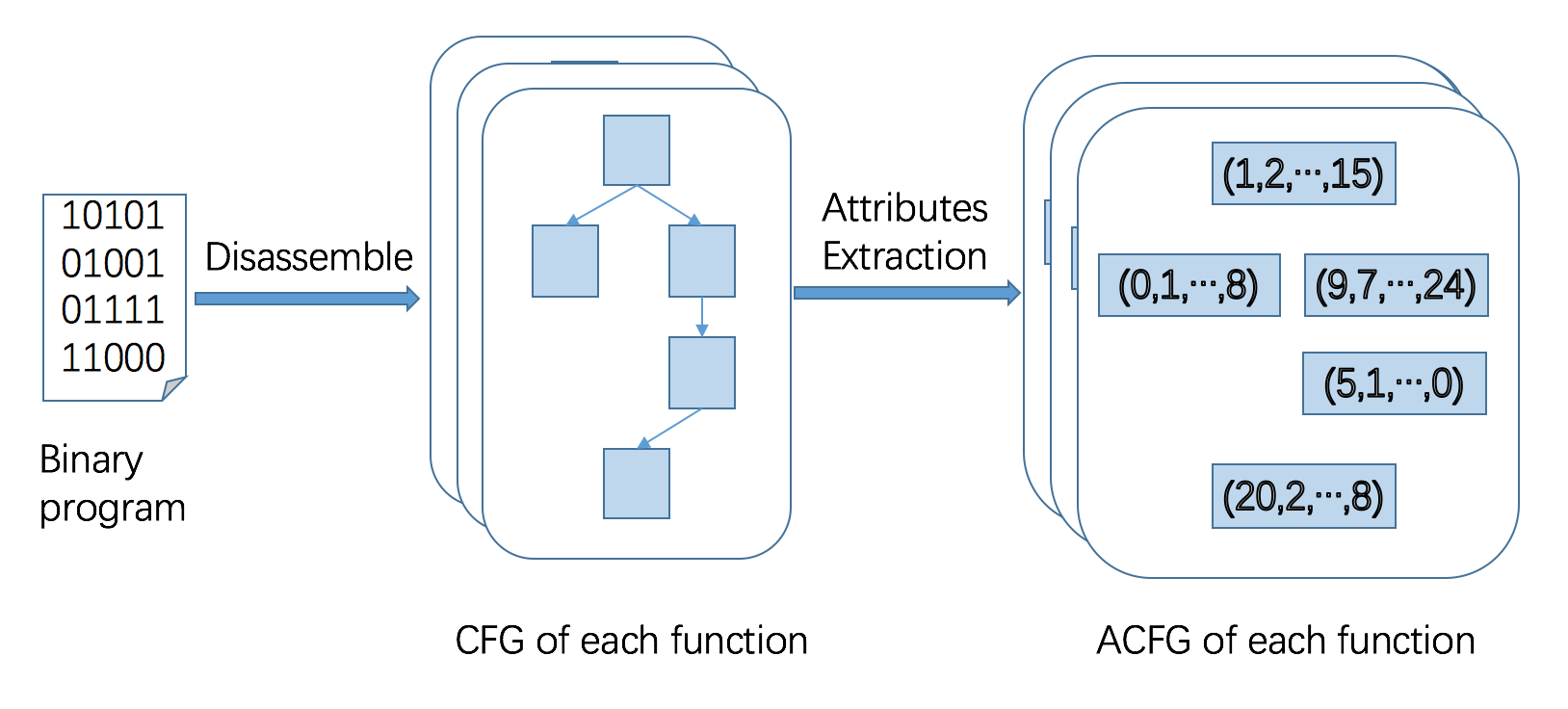}
\caption{The Workflow of data preprocessing.}
\DeclareGraphicsExtensions.
\label{d_p_appendix}
\end{figure}

\begin{table*}[!t]
\centering 
\caption{The used attributes of basic blocks.}
\begin{tabular}{llcccc}
\toprule[1.1pt]
Type & Attributes & Num \\
\midrule[1.1pt]
\multirow{1}{*}{Instructions}
& The num of \emph{call} instruction&\multirow{1}{*}{244}\\
\hline
\multirow{8}{*}{Operand}
& The num of void operand & \multirow{8}{*}{8}\\
& The num of general register operand \\
& The num of direct memory reference operand\\
& The num of operand that consists of a base register or an index register\\
& The num of operand that consists of registers and displacement value\\
& The num of immediate operand\\
& The num of operand that is accessing immediate far addresses\\
& The num of operand that is accessing immediate near addresses\\
\hline
\multirow{3}{*}{Other}
& The num of string "malloc" & \multirow{3}{*}{3}\\
& The num of string "calloc"\\
& The num of string "free"\\
\hline
\multicolumn{2}{c}{All attributes num}& 255\\
\bottomrule[1.1pt]
\label{appendix-full-attributes}
\end{tabular}
\end{table*}

Now,
we show how to vectorize a binary program.
Figure \ref{d_p_appendix} shows the workflow of data preprocessing.
First,
we disassemble the binary program to get the \emph{CFGs} of its functions.
Then,
we extract attributes for basic blocks and transform each basic block into a numerical vector.
The attributes are used to characterize a basic block,
and they can be statistical,
semantic and structural.
Here,
we only extract the statistical attributes for the following reasons.
The first reason is for efficiency.
As indicated in  
\cite{Feng2016Scalable},
the cost of extracting semantic features such as I/O pairs of basic blocks is too expensive.
Second,
the graph embedding network can learn the structural attributes automatically.
We extract 255 attributes in total.
Table \ref{appendix-full-attributes} shows all the 255 attributes,
and all the instruction type-related attributes can be found in Section 5.1 of  
\cite{intel}.

There are mainly three kinds of attributes: 
instruction-related attributes,
operand-related attributes
and string-related attributes.
Then, 
each basic block can be represented by a 255-dimensional vector,
and the binary program now is represented by a set of 255-dimensional vectors.


\subsection{Model Structure}

Based on the discussion in Section 2, 
we choose to adapt 
\emph{Graph Embedding Network}
\cite{Dai2016Discriminative}
\cite{Xu2017Neural}  
as the core of our vulnerability prediction model.
Firstly,
we give a brief introduction of the graph embedding network.
Then,
we detail the design of our model.

\begin{figure}[!t]
\centering
\includegraphics[width=3.5in]{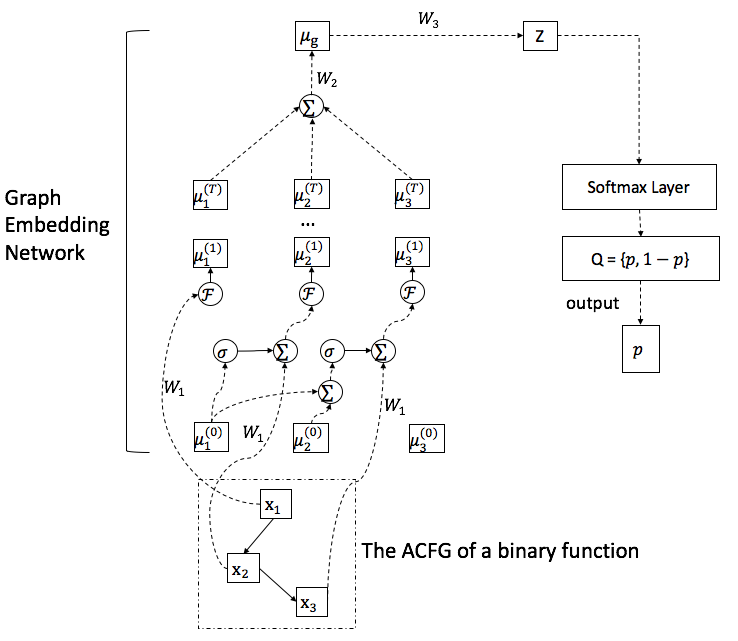}
\caption{The Structure of the vulnerability prediction model. 
$p$ is the output of the model, 
which represents the vulnerable probability of a binary function. }
\DeclareGraphicsExtensions.
\label{network_of_model}
\end{figure}

Graph embedding
is an efficient approach to solve graph-related problems 
\cite{cai2018comprehensive} 
such as node classification,
recommendation and so on.
It transforms a graph into an embedding vector that contains sufficient information of the graph for solving corresponding problems.
In our scenario,
the embedding vectors of a binary function should be able to contain sufficient features for vulnerability prediction.
In addition,
graph embedding can be considered as a mapping $\lambda$, 
which maps a function's \emph{ACFG} $g$ 
into a vector 
$\lambda(g)$.

We leverage a neural network to approximate the mapping $\lambda$,
and train the model with vulnerable and secure binary functions to enable the graph embedding network to learn the features related to vulnerabilities.
As the vulnerability prediction model is required to output the vulnerable probability of a binary function,
we combine the graph embedding network with a pooling layer and a softmax-layer.
The pooling layer transforms the embedding vector into a 2-dimensional vector $Z$,
and the softmax-layer maps the 2-dimension vector $Z$ of arbitrary real values into another 2-dimension vector $Q$,
where the value of each dimension is in the range $[0,1]$.
The first dimension represents the vulnerable probability, 
which is represented by $p$.  
The second dimension represents the secure probability, 
and naturally the value is $1-p$. 
The whole model is trained by labeled data end-to-end, 
and the parameters of the model can be learned by minimizing a loss function.

\begin{table}[!t]
\centering 
\caption{Notations.}
\begin{tabular}{cl}
\toprule[1.1pt] 
Notation & Meaning\\
\midrule[1.1pt]
$a$ & The number of attributes of a basic block\\
$g$ & The ACFG of a binary function\\
$V$ & The set of vertices\\
$E$ & The set of edges\\
$v$  & a vertex  in $V$\\
$x_{v}$ & The attributed vector of vertex $v$\\
$d$ & The dimension of an embedding vector\\
$\mu_{v}$ & The embedding vector of vertex $v$\\
$\mu_{g}$ & The graph embedding vector of ACFG $g$\\
\bottomrule[1.1pt]
\label{appendix-notations}
\end{tabular}
\end{table}

Below is the formalization of the model. 
Table \ref{appendix-notations} shows all the notations related to the model.
and the structure of the model is presented in Figure \ref{network_of_model}.
The input of the model is the \emph{ACFG} $g$ of a binary program function, 
$g=<V,E,\phi>$. 
Each basic block $v \in V$ in \emph{ACFG} has an attribute vector $x_{v}$ which can be constructed according to 
all selected attributes.
The number of attributes for each basic block is $a$.
Thus, 
$x_{v}$ is an $a$-dimensional vector.
For each basic block $v$, 
the graph embedding network computes an embedding vector $\mu_{v}$,
which combines the topology information of the graph. 
The dimension of the embedding vector $\mu_{v}$ is $d$.
Let $N_{v}$ be the set of neighboring vertices of $v$.
Since ACFG is a directed graph, 
$N_{v}$ can be considered as the set of precursor vertices of $v$.
Then the embedding vector $\mu_{v}$ can be computed by
$\mu_{v}=F(x_{v}, \sum_{j \in N_{v}}(\mu_{j}))$,
where
$F$ is a non-linear function that can be $tanh$, 
$sigmoid$ etc. 
The embedding vector $\mu_{v}$ is computed for $T$ iterations. 
For each iteration $t$, 
the temp embedding vector can be get by equation
$\mu_{v}^{(t)}=F(W_{1}x_{v}+\sigma(\sum_{j \in N_{v}}(\mu_{j})^{(t-1)}))$,
where $x_{v}$ is an $a \times1$ vector,
and $W_{1}$ is a $d \times a$ matrix.
The initial embedding vector $\mu_{v}^{(0)}$ is set to zero.  
After $T$ iterations, 
$\forall v \in V$,
we obtain the final graph embedding vector $\mu_{v}^{T}$. 
$\sigma$ is a $n$-layer fully-connected neural network with parameters 
$P={\{P_{1}, P_{2}, ..., P_{n} \}}$. 
Let $ReLU(\cdot)=max\{0,\cdot\}$ be a rectified linear unit.
We have
$\sigma(x)=P_{1}ReLU(P_{2}...ReLU(P_{n}(x)))$.
After $T$ iterations, 
we can get the final graph embedding vector $\mu_{v}^{T}$ for each vertex $v \in V$.
Then the graph embedding vector $\mu_{g}$ of the \emph{ACFG} $g$ can be represented by the summation of the embedding vector of each basic block, 
i.e.,
$\mu_{g}=W_{2}(\sum_{v \in V}(\mu_{v}^{T}))$,
where
$W_{2}$ is a $d \times d$ matrix.
To compute the vulnerable probability of the function,
we map the graph embedding vector into a 2-dimensional vector $Z={\{z_{0},z_{1}\}}$,
i.e.,
\begin{equation}
Z=W_{3}\mu_{g},
\end{equation}
where
$W_{3}$ is a $2 \times d$ matrix.
Then, we use a softmax function to map the values of $Z$ into the vector $Q={\{p, 1-p\}}$, 
$p \in [0,1]$,
i.e.,
\begin{equation}
Q=Softmax(Z)
\label{7}
\end{equation}

The value of $p$ is the output of the model,
which represents the vulnerable probability of the binary program function $g$.

\subsection{Train and Use the Model}
In order to predict the vulnerable probability,
the model needs be trained with labeled data,
where the label is either "vulnerable" or "secure".
For the \emph{ACFG} $g$ of a function, 
the label $l$ of $g$ is 0 or 1,
where $l=0$ means the function has at least one vulnerability, 
and $l=1$ means the function is secure.

The model's training process is similar to a classification model.
Then the parameters of $M$ can be learned by optimizing the following equation:
\begin{equation}
\min \limits_{{W_{1}, W_{2}, W_{3},..., P_{1}, P_{2},...,P_{n}}} \sum_{i=1}^{m}(H(Q,l)),
\label{equation-opt}
\end{equation}
where $m$ is the number of training data,
and $H$ is a cross-entropy loss function. 
We optimize  Equation (\ref{equation-opt}) with a stochastic gradient descent method. 
Thus,
the vulnerability prediction model can be trained in this way.


Although the model's training process is similar to training a classification model,
when using the model,
the classification result that whether a binary function is vulnerable or not is
too coarse-grained and not suitable for the subsequent fuzzing.
Therefore,
we choose to use the vulnerable probability $p$ as the output of the model.

\section{Vulnerability-Oriented Fuzzing}

\begin{figure}[!t]
\centering
\includegraphics[width=3.5in]{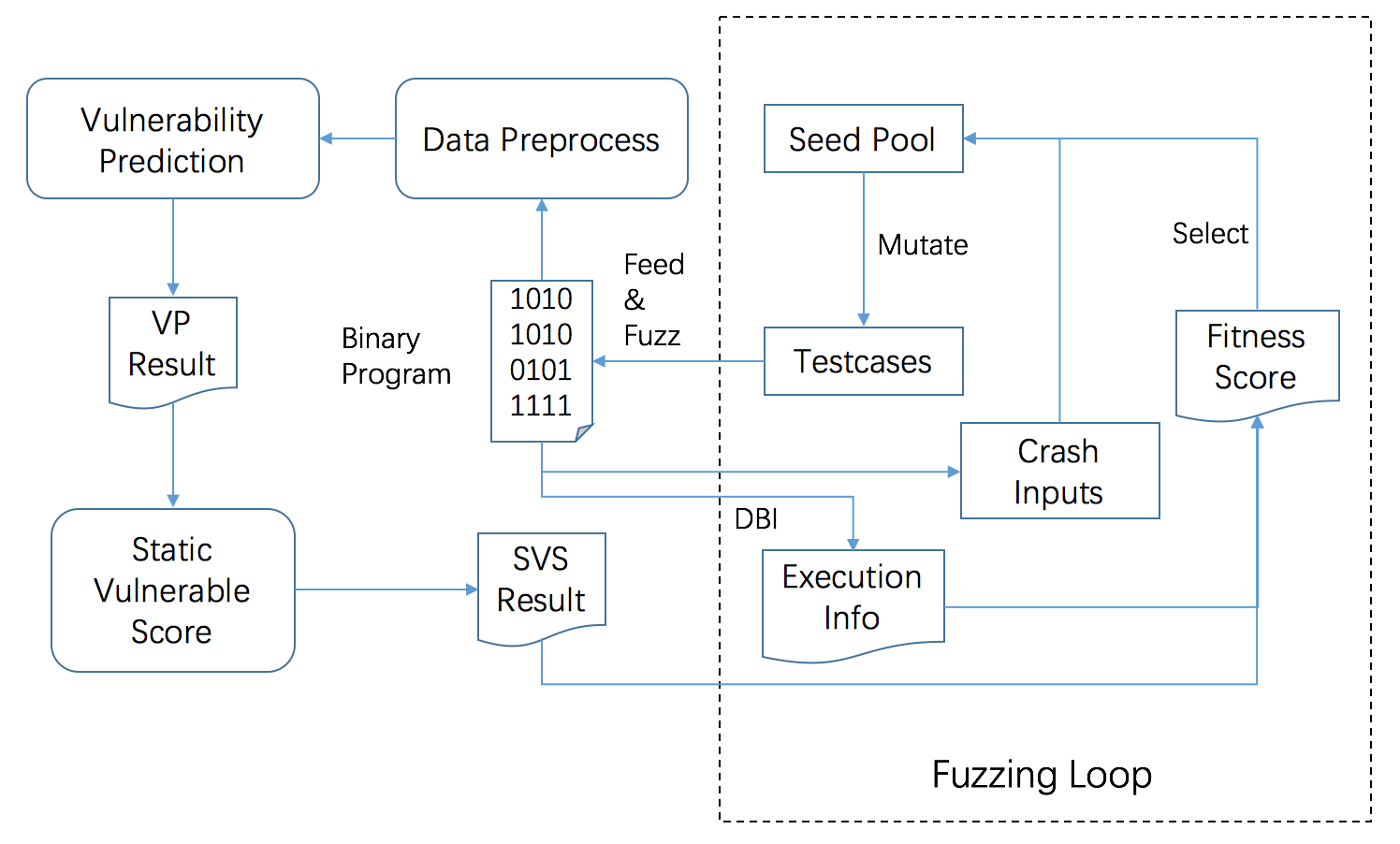}
\caption{Vulnerability-Oriented Fuzzing. DBI: Dynamic Binary Instrumentation,
SVS: Static Vulnerable Score,
and VP: Vulnerability Prediction.}
\DeclareGraphicsExtensions.
\label{fuzzing}
\end{figure}

Based on the result from the prediction model, 
the vulnerability-oriented fuzzer will pay more attention to the functions with high probabilities.
Figure \ref{fuzzing} shows the workflow of vulnerability-oriented fuzzing,
where V-Fuzz leverages an evolutionary algorithm to generate inputs which tend to arrive at the vulnerable components.
Specifically,
for a binary program,
V-Fuzz uses the data process module to disassemble the binary to get the \emph{ACFG} of each function, 
which is the input of the vulnerability prediction model.
Then, 
the prediction model will give each binary program function a \emph{Vulnerability Prediction (VP)}.
Based on the \emph{VP} result, 
each basic block in the program is given a \emph{Static Vulnerable Score (SVS)},
which will be used later to evaluate the executed test cases.

The fuzzing test is a cyclic process.
Like most mutation-based evolutionary fuzzers, 
V-Fuzz maintains a seed pool, 
which is used to save high-quality inputs as seeds.
V-Fuzz starts to execute the binary program with some initial inputs that are provided by users.
Meanwhile,
it uses \emph{Dynamic Binary Instrumentation (DBI)} to track the execution information of the program such as basic block coverage.
Based on \emph{SVS}
and the execution information,
V-Fuzz will calculate a \emph{fitness score} for each executed testcase.
The testcases with high \emph{fitness scores} are considered as high-quality inputs,
and will be sent to the seed pool.
In addition,
the executed testcases which trigger crashes will also be sent to the seed pool,
regardless of their fitness scores.
The detailed method for calculating \emph{fitness score} will be given in Section 5.2.
Next,
V-Fuzz generates the next generation testcases by mutating the seeds in the seed pool.
In this way,
V-Fuzz continues to execute the program with new generated inputs until the end conditions are met.
Below, 
we elaborate the workflow of vulnerability-oriented fuzzing.

\subsection{Static Vulnerable Score}

Based on the \emph{VP} result,
V-Fuzz gives each basic block a \emph{Static Vulnerable Score (SVS)}.
For a function $f$, 
we assume its vulnerable probability is $p_{v}$, 
and it has $\iota$ basic blocks $f={{b_{1},b_{2},...,b_{\iota}}}$. 
For $b_{i} \in f$,
$b_{i}$'s \emph{SVS}, 
denoted by $SVS(b_{i})$,
can be calculated by the following equation:
\begin{equation}
SVS(b_{i})=\kappa*p_{v}+\omega,
\end{equation}
where $\kappa$ and $\omega$ are constant parameters that should be obtained from fuzzing experiments.
Hence,
the basic blocks that belong to the same function have the same $SVS$ values.
For parameter $\kappa$,
we fuzz 64 Linux programs (e.g., some programs in binutils),
and 20 of them have crashes.
Then,
we fuzz these 20 programs individually with the value of $\kappa \in [10,100]$,
and we observe that when $\kappa \in [15,25]$,
the fuzzing test performs the best.
Therefore,
we set $\kappa=20$ as the default value.
For parameter $\omega$,
it is used to avoid $SVS=0$ when functions have very low vulnerable probabilities .
As if $SVS(b_{i})=0$,
it represents that $b_{i}$ has no meaning for fuzzing and becomes trivial,
which is against our design principle of V-Fuzz.
Therefore,
we set the value of $\omega=0.1$,
which is small and can make $SVS>0$ all the time.

Based on the approach of calculating $SVS$,
V-Fuzz assigns more weight to the basic blocks that are more likely to be vulnerable,
which will further assist the fuzzer to generate inputs that are more likely to cover the basic blocks.

\subsection{Seed Selection Strategy}

Algorithm \ref{alg-seed}
shows the seed selection strategy of V-Fuzz.
Specifically,
V-Fuzz leverages an evolutionary algorithm to select seeds which are more likely to arrive at the vulnerable components.

\begin{figure}[!t]
\centering
\includegraphics[width=3.5in]{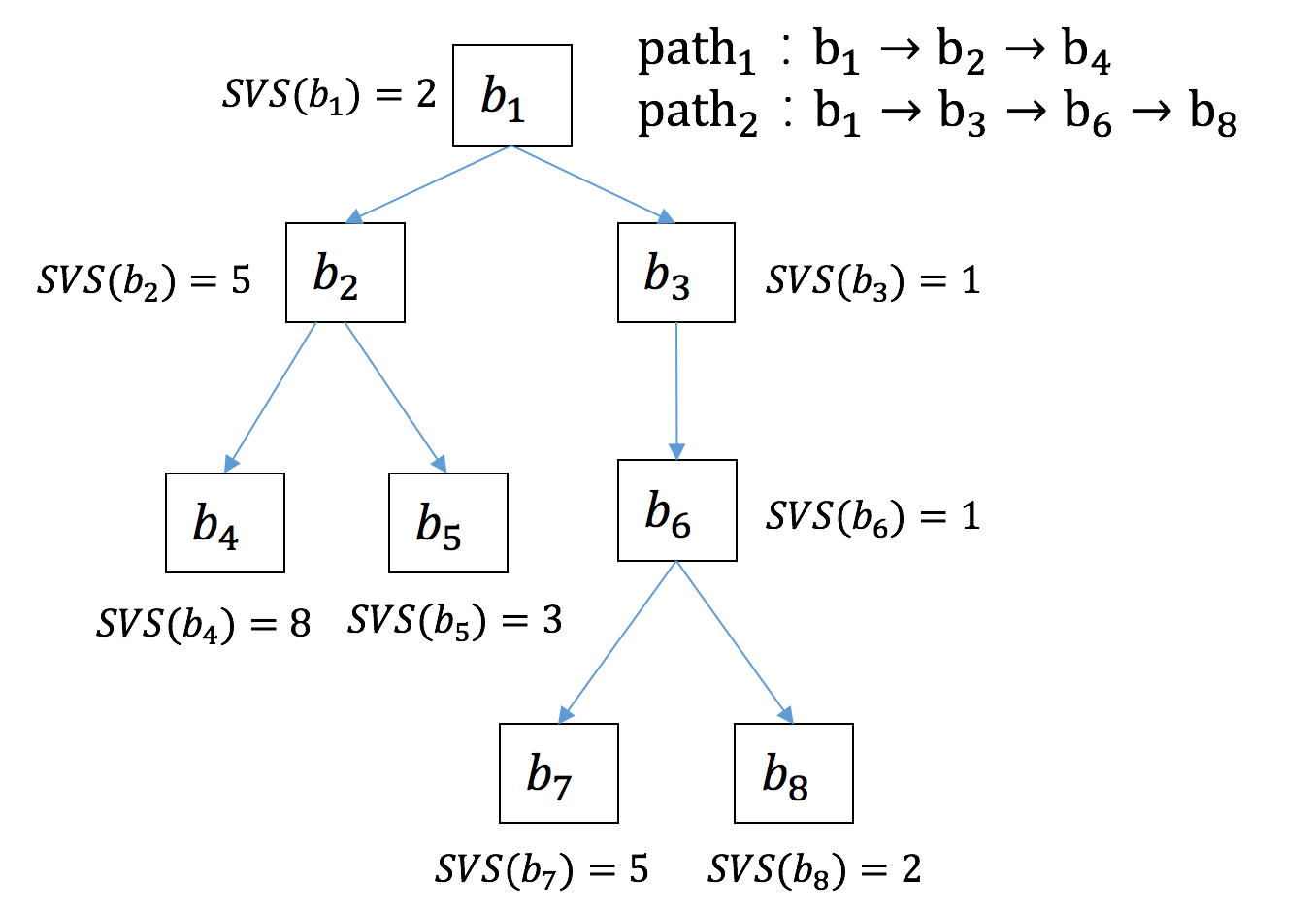}
\caption{An example for \emph{fitness score} calculation.}
\DeclareGraphicsExtensions.
\label{fitness}
\end{figure}

After giving every basic block a $SVS$, 
V-Fuzz enters the fuzzing loop.
During each loop, 
V-Fuzz monitors the program to check if it has exceptions such as crashes.
If the input causes a crash,
then the input is added to the seed pool.
Once an execution has completed, 
V-Fuzz records the execution path for the input. 
The \emph{fitness score} of the input is the sum of the 
\emph{SVS} values 
of the basic blocks that are on the execution path. 
Figure \ref{fitness} shows an example for \emph{fitness score} calculation.
We assume there are two inputs $i_{1}$,
$i_{2}$ in this generation.
The execution paths of the two inputs are $path_{1}$ and $path_{2}$ respectively.
Suppose 
$path_{1}$ is $b_{1} \rightarrow b_{2} \rightarrow b_{4}$,
and 
$path_{2}$ is $b_{1} \rightarrow b_{3} \rightarrow b_{6} \rightarrow b_{8}$.
The \emph{fitness score} of input $i_{1}$ and $i_{2}$ are $f_{1}$ and $f_{2}$
respectively.
Then,
$f_{1}=2+5+8=15$,
$f_{2}=2+1+1+2=6$.
As $f_{1}$ is larger than $f_{2}$,
the input $i_{1}$ will be selected as a seed.
It should be noted that,
if any input causes a crash,
no matter how low the \emph{fitness score} it has,
it will be sent to the seed pool. 

In this way,
V-Fuzz not only utilizes the information of the vulnerability prediction model,
but also considers the actual situation.
Therefore,
V-Fuzz can mitigate the potential weakness of the vulnerability prediction model.

\begin{algorithm}
\small
\algsetup{linenosize=\tiny}
	\caption{Seed Selection Algorithm}\label{alg-seed}
	\begin{algorithmic}
	\algsetup{linenosize=\tiny}
	\REQUIRE
	Binary Program: $p$;
	The set of all basic blocks of $p$: $B$;
	The set of initial inputs: $I$;
	The set of seed pool: $S$;
	The set of testcases: $T$;
	\STATE $T=I$ 
	\WHILE {In the fuzzing loop}
	\FOR  {$t$ in $T$}
	\STATE $Path, Ret=EXE(p,t)$ 
	\STATE {$fitness(t)=\sum_{b \in Path}(SVS(b))$}
	\IF {$Ret(i) \in CRASH$}
	\STATE $S.add(t)$
	\ENDIF
	\ENDFOR
	\STATE $Q=SelectKfitness(I)$ 
	\STATE $S.add(Q)$
	\STATE $T=Mutate(S)$
	\ENDWHILE
\end{algorithmic}  
\end{algorithm}

\subsection{Mutation Strategy}

V-Fuzz is a mutation-based evolutionary fuzzing system,
which generates new inputs by mutating the seeds.
Like most of mutation-based fuzzers,
the mutation operations are bit/byte flips,
inserting "interesting" bits/bytes,
changing some bits/bytes,
selecting some bytes from several seeds and splicing them together and so on.

\begin{figure}[!t]
\centering
\includegraphics[width=3.5in]{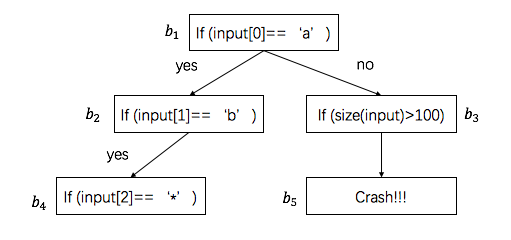}
\caption{A simple \emph{CFG}.}
\DeclareGraphicsExtensions.
\label{cw}
\end{figure}

The design of the mutation strategy is very important,
as an appropriate strategy can help the fuzzer generate good inputs which can find new paths or crashes.
For example,
Figure \ref{cw} gives the \emph{CFG} of a simple program.
Assume that there is a seed string $s_{1}=``abx"$,
which can cover the basic block $b_{1}$, $b_{2}$.
Another seed string $s_{2}=``qwerty"$ covers no basic block of the program.
It is obvious that the new inputs mutated from $s_{1}$ are more likely to cover new basic blocks than those mutated from $s_{2}$.

It is worth noting that if we want to get "ab*" by mutating "abx",
the mutation must be slight,
which only changes a small part of the original seed.
However,
if the fuzzer has spent too much time doing the slight mutation operations, 
while making no progress,
the fuzzer should change its mutation strategy and pay more attention to other paths.
In this example,
if the fuzzer gets "stuck" for a long time by performing the slight mutation on "abx",
it would be better to choose heavy mutation that changes more about the original seed,
which may help the fuzzer find the basic block $b_{3}$.
Therefore,
the fuzzer should dynamically adjust its mutation strategy according to the actual fuzzing states.

We classify the mutation strategies into slight mutation and heavy mutation.
In order to help the fuzzer determine the selection of the mutation strategy,
we define \emph{Crash Window (CW)},
which is a threshold to determine the selection of mutation strategy.
Consequently,
we assume that the number of generations whose inputs have not found any new path or crash is denoted by notation $\zeta$.
If $\zeta>CW$,
the fuzzer should select heavy mutation strategy.
Furthermore,
we propose the \emph{Crash Window Jump Algorithm}
to adjust the value of \emph{CW} optimally.

The main idea of the \emph{Crash Window Jump Algorithm}
is as follows:
First,
we assume that the initial value of \emph{CW} is $ini\_cw$,
its maximum value is $max\_cw$ and its minimum value is $min\_cw$.
The value of \emph{CW} starts from $ini\_cw$,
and the fuzzer selects slight mutation as its initial mutation strategy.
During the fuzzing process,
if $\zeta>CW$,
the fuzzer will change its mutation strategy to heavy mutation,
and will double the value of \emph{CW}.
Once an input finds a new path or a crash,
then we set $\zeta=0$ and the new value of \emph{CW} as the half of its former value.
Algorithm \ref{appendix-cwja} shows the pseudo-code of the \emph{Crash Window Jump Algorithm}.

\begin{algorithm}
\caption{Crash Window Jump Algorithm}
\label{appendix-cwja}
\begin{algorithmic}
\small
  \REQUIRE
  Binary Program: $p$;
  The initial crash window: $ini\_cw$;
  Max Crash Window: $max\_cw$;
  Min Crash Window: $min\_cw$;
  The current Crash Window: $CW$;
  The set of seed pool: $S$;
  The set of testcases: $T$;
  The mutation strategy: $MS$;
  $no\_crash=True$;
  $no\_new\_bb=True$;
  \STATE $CW=ini\_cw$
  \STATE $T=I$ 
  \STATE $\zeta=0$
  \STATE $MS=slight\_mutate$
  \WHILE {In the fuzzing loop}
    \FOR  {$t$ in $T$}
              \STATE $EXE(p,t)$ 
      \IF {find crash}
      \STATE $no\_crash=False$
      \ENDIF
      \IF{find new basic block}
      \STATE $no\_new\_bb=False$
      \ENDIF
    \ENDFOR
    
    \IF {$no\_crash==True  \&\&   no\_new\_bb==True$}
    \STATE $\zeta++$
      \IF {$\zeta>CW$}
      \STATE $MS=heavy\_mutate$
        \IF {$CW >=min\_cw \times 2$}
        \STATE $CW=CW/2$
        \ENDIF
      \ENDIF
    \ELSE
      \STATE $\zeta=0$
      \STATE $MS=slight\_mutate$
      \IF {$CW<= max\_cw/2$}
      \STATE $CW=CW \times 2$
      \ENDIF
    \ENDIF
    \STATE $S=select\_good\_seed(T)$
    \STATE $T=MS(S)$
  \ENDWHILE
\end{algorithmic}  
\end{algorithm}

\section{Implementation and Settings}

\subsection{Vulnerability Prediction}
The vulnerability prediction module consists of two main components: 
the \emph{ACFG} extractor and the vulnerability prediction model.
For the \emph{ACFG} extractor,
we implement it by writing a plug-in on the famous disassembly tool IDA Pro  
\cite{idapro}.
For the vulnerability prediction model,
we implement it based on PyTorch 
\cite{pytorch},
which is a popular deep learning framework.

We train the vulnerability prediction model on a server which is equipped with 
two Intel Xeon E5-2640v4 CPUs (40 cores in total) running at 2.40GHz,
4 TB HDD,
64 GB memory,
and
one GeForce GTX 1080 TI GPU card.

\subsection{Vulnerability-Oriented Fuzzing}
For vulnerability-oriented fuzzing, 
we implement the fuzzer based on VUzzer
\cite{Rawat2017VUzzer}, 
an application-aware evolutionary fuzzer which focuses on fuzzing binary programs.

We conduct the fuzzing test on a virtual machine with Ubuntu 14.04 LTS.
The virtual machine is configured with 32-bit single core 4.2GHz Intel CPU and 4 GB RAM,
During the fuzzing test,
we observe that the fuzzing process takes less than 1GB of memory.

\section{Evaluation}
In this section,
we evaluate the performance of V-Fuzz.
Since V-Fuzz consists of two components,
we will present the evaluation results in two parts:
the vulnerability prediction and the vulnerability-oriented fuzzing.

\subsection{Vulnerability Prediction Evaluation}

\subsubsection{Data Selection}

The dataset used for training and testing the vulnerability prediction model is published by the National Institute of Standards and Technology (NIST)
\cite{NIST},
as this dataset has been widely used in many vulnerability related work 
\cite{Li2018VulDeePecker}
\cite{Han2018Enhancing}.
We use the codes of Juliet Test Suite v1.3  
\cite{juliet} as our training and testing data,
which is a collection of test cases in the C/C++ language,
and each function in this dataset is labeled with "good" or "bad".
A function labeled with "good" means it does not have flaws,
while one labeled with "bad" means it has at least one flaw.
Each "bad" example has a Common Weakness Enumeration IDentifier (CWE ID).
Juliet Test Suite v1.3 has examples of 118 different CWEs in total.
As fuzzing is suitable for discovering bugs related to memory,
we select some CWE samples which are related to memory errors from Juliet Test Suite v1.3,
which are shown in Table \ref{CWE}.

As Table \ref{CWE} shows,
we collect 111,540 labeled function samples in total,
which include 78,511 secure samples and 33,029 vulnerable samples.
The top three types of CWEs are 
\texttt{Integer Overflow}   
(the number is 26,982 ), 
\texttt{Heap Based Buffer Overflow}
(the number is 18,522)
and
\texttt{Stack Based Buffer Overflow}
(the number is 15,403).
These CWE testcases take almost half of all the data 
(54.6\%).
In addition,
these 3 types of vulnerabilities are the most common ones in real-world.
Thus,
the distribution of our selected datasets is similar as the distribution of  real-world vulnerabilities.

From Table \ref{CWE},
we randomly select 40,000 samples as the training data \texttt{TRAIN-DATA}.
Then,
we again randomly select 4,000 samples as the testing data \texttt{TEST-DATA},
which has no overlap with \texttt{TRAIN-DATA}.
Table \ref{dataset} presents the information of the 3 datasets.

\begin{table}[!t]
\centering 
\footnotesize
\caption{The Types of CWE.}
\setlength{\tabcolsep}{1mm}{
\begin{tabular}{clcccccc}
\toprule[1pt]
CWE & Type & \#Secure & \#Vulnerable & Total \\
\midrule[1pt]
121 & Stack Based Buffer Overflow & 10,187 & 5,216 & 15,403\\
122 & Heap Based Buffer Overflow & 12,263 & 6,259 & 18,522\\
124 & Buffer Under write & 4,183 & 2,031 & 6,214\\
126 & Buffer Over Read &3,019 & 1,376 & 4,395\\
127 & Buffer Under Read &4,183 &2,031 & 6,214\\
134 & Uncontrolled Format String & 6,833 &  2,357 & 9,190\\
190 & Integer Overflow & 20,187& 6,795 & 26,982\\
401 & Memory Leak  & 6,756 & 2,303 & 9,059\\
415 & Double Free & 4,204 & 1,448 & 5,652\\
416 & Use After Free & 1,760 & 470 & 2,230\\
590 & Free Memory Not On The Heap & 4,049 & 2,250 &6,299 \\
761 & Free Pointer Not At Start & 887 & 493 &1,380\\
\hline
\multicolumn{2}{c}{Total} & 78,511 & 33,029 &111,540 \\
\bottomrule[1pt]
\label{CWE}
\end{tabular}}
\end{table}

\begin{table}[!t]
\setlength{\tabcolsep}{0.6mm}{
\centering 
\footnotesize
\caption{Datasets.}
\begin{tabular}{lllll}
\toprule[1.1pt]
Dataset & \#Vulnerable & \#Secure & Total  &  Remarks \\ 
\midrule[1.1pt]
ALL-DATA  &  78,511  &  33,029  & 111,540  & All data in Table 2 \\
TRAIN-DATA  & 20,000 & 20,000 & 40,000 & Selected from ALL-DATA\\
TEST-DATA & 2,000 & 2,000 & 4,000  & Selected from ALL-DATA\\
\bottomrule[1.1pt]
\label{dataset}
\end{tabular}}
\end{table}

\subsubsection{Pre Experiments}

First,
we conduct some pre-training experiments to determine the default parameters of the model.
We use stochastic gradient descent as the optimization algorithm
and set the learning rate equal to 0.0001.
Based on the results of the pre-training experiments,
we set the depth of the network as 5,
the embedding size as 256,
and the number of iterations as 3.
We take the above setting as the default of our model.
Then,
we use \texttt{TRAIN-DATA} to train the vulnerability prediction model, 
and use \texttt{TEST-DATA} to test the model.

\subsubsection{Evaluation Metrics}

We use three metrics to evaluate the performance of the model:
accuracy, 
recall and loss.
The first two metrics reflect the model's capability of predicting vulnerable functions.
Additionally,
we evaluate whether the model converges by observing  the value of loss.
Next,
we show the method to calculate these three metrics and show the performance of the model using these metrics.

Suppose the number of testing samples is $\hat{L}$.
Among these samples,
the number of samples labeled as "vulnerable" is $\hat{V}$,
and the number of samples labeled as "secure" is $\hat{L}-\hat{V}$.
Based on the prediction outputs,
we sort the testing examples in descending order of their predicted values (i.e., vulnerable probability).
Then,
we select the \emph{top-K} testing samples.
Assuming among the \emph{top-K} samples,
the number of samples labeled as "vulnerable" is $\hat{v}$,
and the number of samples labeled as "secure" is $K-\hat{v}$.
Thus,
we can calculate the accuracy of the \emph{top-K} testing data as $accuracy=\hat{v}/K$.
When the threshold \emph{K} equals to the number of $\hat{V}$,
we can calculate the recall as $recall=\hat{v}/\hat{V}, (K=\hat{V})$.
Finally,
we leverage the cross-entropy loss function to calculate loss.

\begin{figure*}[!t]
\centering
	\begin{minipage}{0.33\linewidth}
	\centering
	\includegraphics[width=5.8cm, height=3cm]{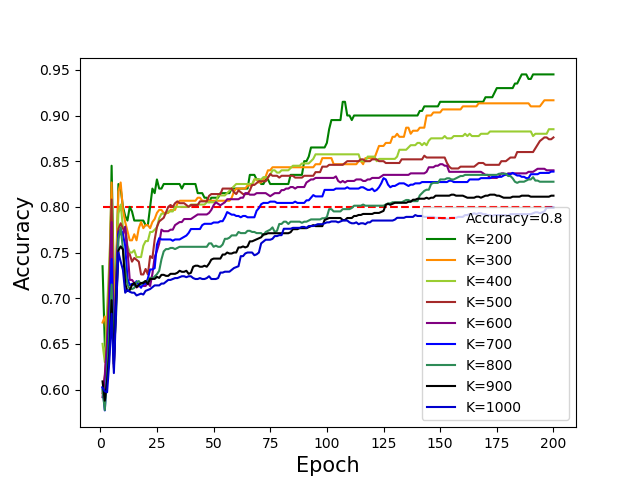}
	\subcaption{Accuracy.}
	\label{perf-a}
	\end{minipage}
	\begin{minipage}{0.33\linewidth}
	\centering
	\includegraphics[width=5.8cm, height=3cm]{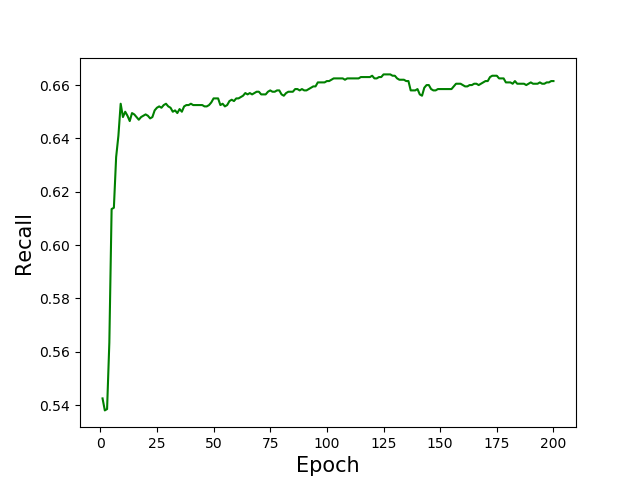}
	\subcaption{Recall.}
	\label{perf-b}
	\end{minipage}
	\begin{minipage}{0.33\linewidth}
	\centering
	\includegraphics[width=5.8cm, height=3cm]{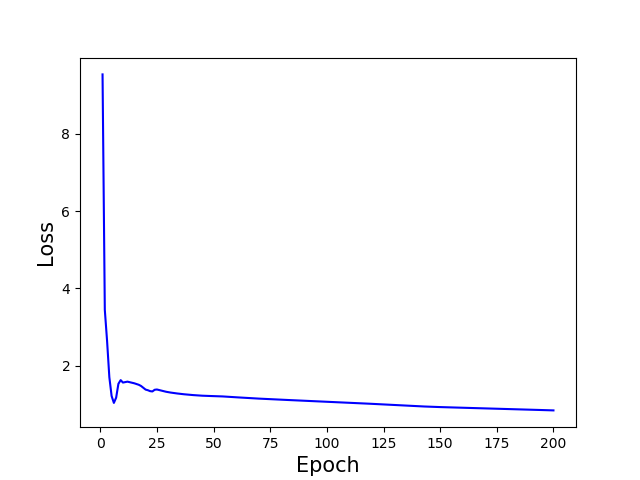}
	\subcaption{Loss.}
	\label{perf-c}
	\end{minipage}
\caption{The performance of the model.}
\label{perf}
\end{figure*}

Figure \ref{perf}
shows the performance of the model evaluated by the three metrics.
Figure \ref{perf-a} presents the prediction accuracy when setting the threshold $K$ to different values 
(in range $[200,1000]$),
from which we can observe that the accuracy of the model is high
(greater than 80\%).
Figure \ref{perf-b} presents the prediction recall.
From the figure,
we can observe that the prediction recall rises from 53\% to 66\% during the training process.
Its value increases very soon and becomes stable within 10 training epochs.
Figure \ref{perf-c} presents the value of loss,
from which we can observe that its value drops very quickly in the first 6 epochs,
and remains stable after that.

From the above results,
we can see that this model is capable for vulnerable function prediction.

\subsubsection{Hyperparameters Analysis}

\begin{figure*}[!t]
  \centering
	\begin{minipage}{0.33\linewidth}
	\centering
	\includegraphics[width=5.8cm, height=3cm]{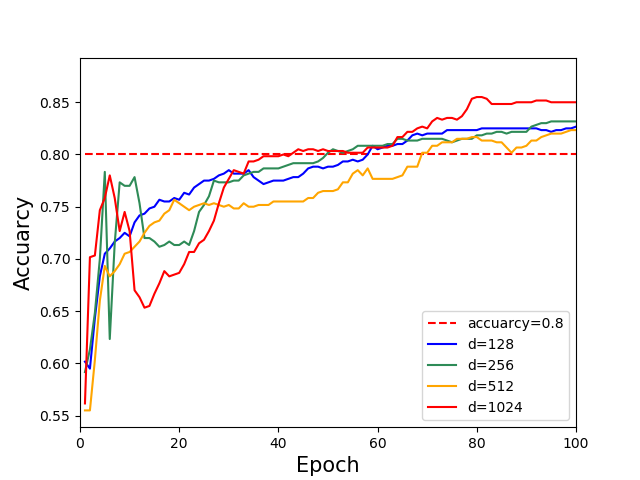}
	\subcaption{Accuracy of different embedding size $d$.}
	\label{hyper-a}
	\end{minipage}
	\begin{minipage}{0.33\linewidth}
	\centering
	\includegraphics[width=5.8cm, height=3cm]{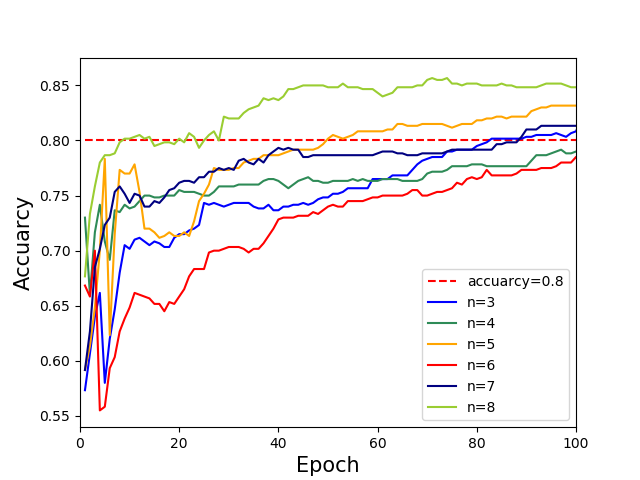}
	\subcaption{Accuracy of different depth $n$.}
	\label{hyper-b}
	\end{minipage}
	\begin{minipage}{0.33\linewidth}
	\centering
	\includegraphics[width=5.8cm, height=3cm]{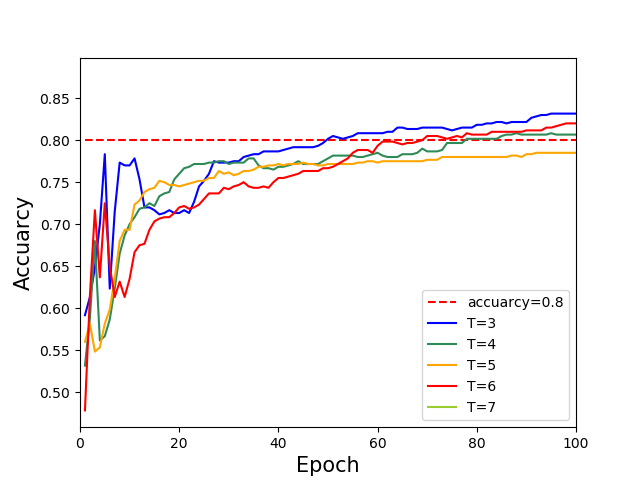}
	\subcaption{Accuracy of different iteration $T$.}
	\label{hyper-c}
	\end{minipage}
\caption{Impact of different hyperparameters.}
\label{hyper}
\end{figure*}

In this part, 
we give an analysis on the hyperparameters.
They are the number of depth,
embedding size and the number of iterations.
Figure \ref{hyper} 
shows the impact of different hyperparameters.
We select the accuracy of \emph{top-K}
(K=600) as the impact indicator to show the effectiveness of different hyperparameters.

\textbf{Embedding Size:}
Figure \ref{hyper-a} shows the impact of different embedding size.
Here,
we test the impact by setting the embedding size as 128, 256, 512 and 1024 respectively.
From the figure, 
we can observe that when the embedding size is 1024,
the accuracy reaches the largest,
and the accuracy of other embedding size is almost the same.
However,
as the overhead will be increased when using a large embedding size,
to balance the overhead and performance,
we choose to select the embedding size as 256 instead of 1024.

\textbf{Depth:}
Figure \ref{hyper-b} shows the impact of different depth.
Here,
we test the depth at 3, 4, 5, 6, 7
and 8.
From the figure,
we can observe that when the depth is 8,
the accuracy reaches the largest.
However,
the accuracy does not have a strictly positive correlation with the number of depth.
For instance, 
the accuracy is the lowest when the depth is 6.
Since training a neural network model with more layers may take more time and computing resource,
we choose to set the depth of our model as 5,
which also has a high accuracy.

\textbf{Iterations:}
The number of iterations $T$ means that the embedding vectors of one vertex contains the information of its \emph{T-hop} neighborhood vertices.
Figure \ref{hyper-c} shows the impact of different iterations.
We test the model with $T \in [3,7]$.
From the figure,
we can observe that when $T=3$,
the accuracy is the highest.
We conjecture the reason may be that the size of our test samples is not very big.
For a basic block,
the information of its \emph{3-hop} neighborhoods is enough for the model to learn the features of the graph.
Therefore,
we choose to set  $T=3$ in our model.
It is worth noting that the number of iterations can be adjusted to make it appropriate for specific training or testing datasets.

\subsubsection{Efficiency}

We evaluate the efficiency of the vulnerability prediction from two aspects:
(1)
\emph{ACFG} extraction time
and
(2)
training time with different model structures.

\begin{figure}[!t]
  \centering
	\begin{minipage}{0.49\linewidth}
	\centering
	\includegraphics[width=4.2cm, height=3cm]{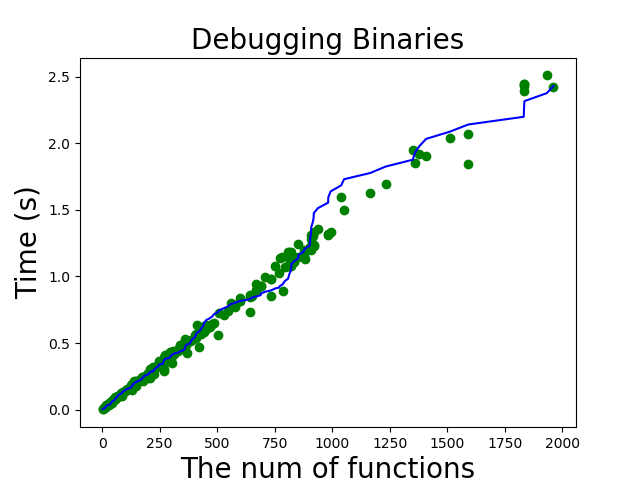}
	\label{acfg_time-a}
	\end{minipage}
	\begin{minipage}{0.49\linewidth}
	\centering
	\includegraphics[width=4.2cm, height=3cm]{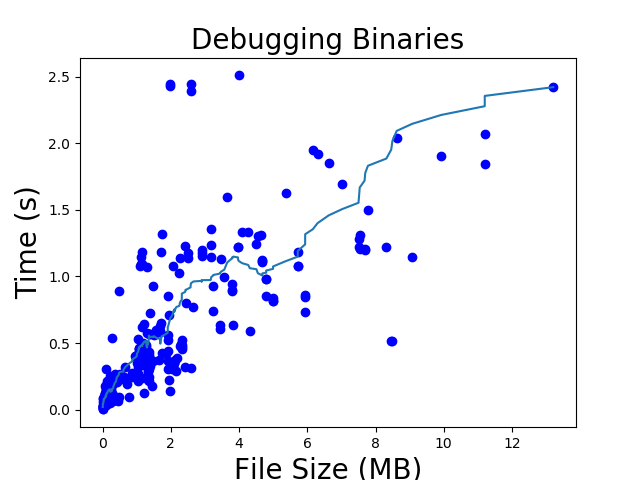}
	\label{acfg_time-b}
	\end{minipage}
	\begin{minipage}{0.49\linewidth}
	\centering
	\includegraphics[width=4.2cm, height=3cm]{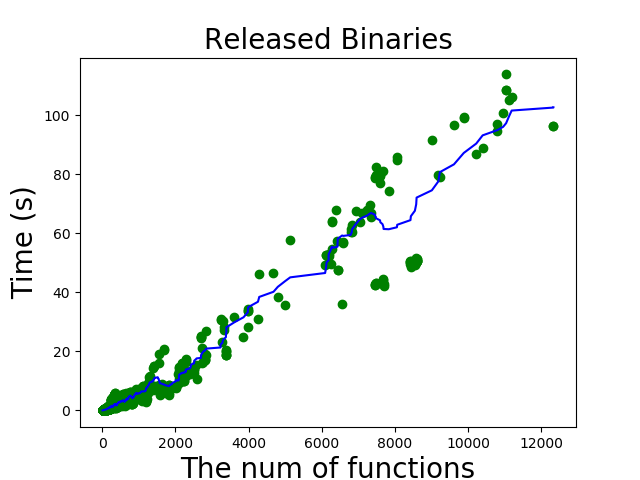}
	\subcaption{The number of functions.}
	\label{acfg_tim-c}
	\end{minipage}
	\begin{minipage}{0.49\linewidth}
	\centering
	\includegraphics[width=4.2cm, height=3cm]{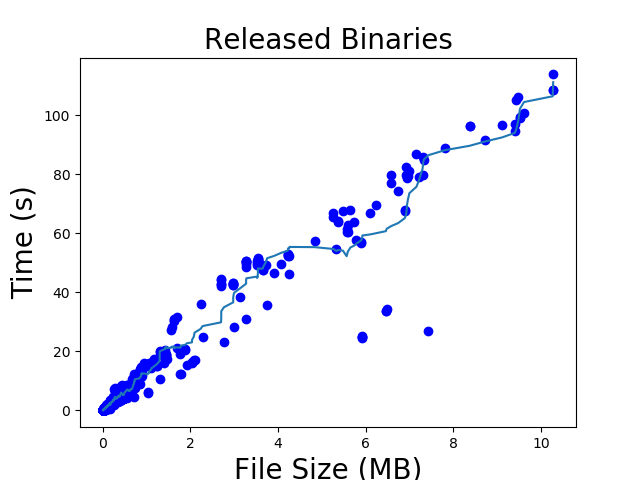}
	\subcaption{File Size.}
	\end{minipage}
\caption{\emph{ACFG} extraction time.}
\label{acfg_time}
\end{figure}

\textbf{\emph{ACFG} Extraction Time.}
We evaluate the \emph{ACFG} extraction time from two aspects:
the number of functions in a binary and the size of the binary.
We collect a lot of binaries samples and test the \emph{ACFG} extraction time of them.
Figure ~\ref{acfg_time} shows the \emph{ACFG} extraction time with different number of functions and file sizes.
Moreover,
we test the debugging binaries and the released binaries respectively.
From the figure,
we have the following observations.
(1) 
The extraction time of debugging binaries is much less than the released binaries.
Since the debugging binaries have more information such as symbolic tables,
which can help improve the speed of disassemble analysis,
while the released binaries don not have so much information and may cost more time to do the disassemble analysis.
(2) 
The extraction time has a positive linear correlation with the number of functions.
(3) 
The extraction time has a positive linear correlation with the value of file size.
(4)
The \emph{ACFG} extraction time is pretty short.
For most of the debugging binaries, 
the extraction time is within 2.5 seconds. 
For most of the released binaries,
the extraction time is within 100 seconds.
Thus,
we can extract the \emph{ACFG} of a binary program efficiently.

\begin{figure}[!t]
\centering
\includegraphics[width=3.5in]{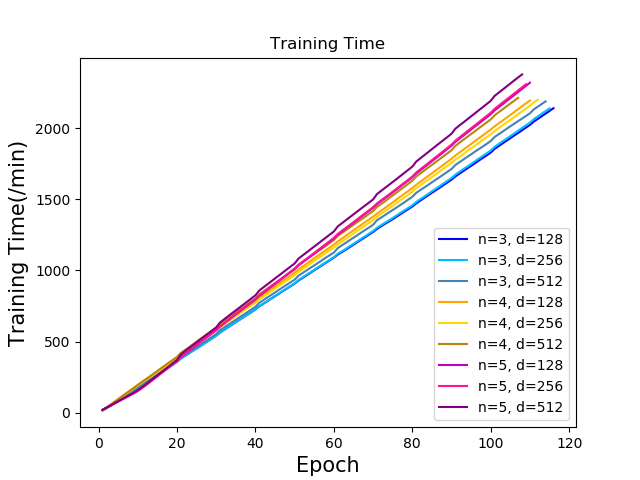}
\caption{Training time with different model parameters: depth $n$ and embedding size $d$.}
\DeclareGraphicsExtensions.
\label{train_time}
\end{figure}

\textbf{Training Time.}
Figure \ref{train_time}
shows the training time with different parameters.
For this model,
the parameters which affect the training time much are depth and embedding size.
From this figure,
we can observe that training time has a positive correlation with both the depth and the embedding size.
Table \ref{table_train_time} 
presents the statistical results of the training time for 50 training epochs with different depth and embedding size.
From the table,
we can see that the average time for training the model in 1 epoch is about 20 minutes.
As the model can converge within 10 epochs,
we only need about 200 minutes to train a valid vulnerability prediction model,
which is significantly more efficient.
In addition,
as the vulnerability prediction model can be trained offline,
it does not affect the time overhead of fuzzing.

\begin{table}[!t]
\centering
\small
\caption{The training time with different parameters.}
\begin{tabular}{cccccccc}
\toprule[1.1pt]
\multirow{2}{*}{Depth} & \multirow{2}{*}{Embedding Size }&\multicolumn{2}{c}{Training Time(min)} \\
 & &50 epochs & average for 1 epoch\\
\midrule[1.1pt]
\multirow{3}{*}{3}&128 &907.70 & 18.15\\
&256 &909.14 & 18.18\\
&512 &934.99 & 18.69\\
\hline
\multirow{3}{*}{4}&128 &984.99 & 19.69\\
&256 & 988.98 &19.77\\
&512 & 1,009.09  & 20.18\\
\hline
\multirow{3}{*}{5}&128 &1,010.24 &20.20 \\
&256 &1,013.58 & 20.27 \\
&512 &1,049.14 & 20.98 \\
\bottomrule[1.1pt]
\label{table_train_time}
\end{tabular}
\end{table}

\subsection{The Evaluation of Fuzzing}

In this section,
we evaluate the performance of V-Fuzz in fuzzing test.
Towards this,
we conduct a number of fuzzing test on 13 different applications as shown in Table \ref{uc}.
These applications are popular real-world open source Linux applications and fuzzing benchmarks.
The real-world Linux applications includes
the audio processing softwares
(e.g., \texttt{MP3GAIN}),
pdf transformation tools
(e.g., \texttt{pdftotext})
and XPS documents library 
(e.g, \texttt{libgxps}).
The fuzzing benchmarks are three programs (\texttt{uniq}, \texttt{base64} and \texttt{who}) of LAVA-M 
\cite{Dolangavitt2016LAVA}.
The reasons for selecting these applications are as follows:
(1) The selected real-world Linux applications are very popular and have been used widely.
(2) As these applications are open source.
the security of these applications may have significant impacts on other applications that are developed based on them.
(3) LAVA-M 
\cite{Dolangavitt2016LAVA} is a famous benchmark for evaluating the performance of vulnerability detection tools.
Many the advanced fuzzing studies test their tools 
\cite{Rawat2017VUzzer}
\cite{CollAFL}
\cite{Chen2018Angora}
with the programs of LAVA-M (a set of programs in LAVA-M).
Even more,
we compare V-Fuzz with several state-of-the-art fuzzers:
VUzzer  
\cite{Rawat2017VUzzer}
AFL  
\cite{AFL}
and AFLFast  
\cite{Pham2016Coverage}.

It should be emphasized that all of the fuzzing experiments are conducted based on the following principles:
(1) All the running environments are the same.
Every single fuzzing test is conducted on a virtual machine equipped with 32-bit single core 4.2 GHz Intel CPU and 4 GB RAM, 
on Ubuntu 14.04 LTS system.
(2) All the initial inputs for fuzzing are same.
(3) The running time of all the fuzzing evaluation is the same.
Therefore,
all the fuzzing experiments are fair and convincing.

There are two main aspects to evaluate a fuzzer's capability of finding bugs: 
\emph{unique crashes},
and \emph{identified vulnerabilities}.
We will demonstrate the performance of V-Fuzz from the two aspects firstly.
Moreover,
we will evaluate the code coverage.

\subsubsection{Unique Crashes}

The capability of finding unique crashes is an important factor to evaluate a fuzzer's performance.
Although a unique crash is not necessarily a vulnerability,
in most cases,
if a fuzzer can find more crashes, 
it can find more vulnerabilities.
More specifically,
a fuzzer with a better capability of finding unique crashes usually has the following characteristics:
(1)
In a limited time, 
it can find more unique crashes.
(2)
For finding a fixed number of unique crashes,
it can find them fast.

Thus, 
we will demonstrate V-Fuzz's performance in finding unique crashes by answering the following two questions.

\textbf{Whether V-Fuzz can find more unique crashes in a limited time?}
We fuzz the 13 programs for 24 hours and compare V-Fuzz with 3 state-of-the-arts fuzzers: VUzzer,
AFL and AFLFast.
In detail, 
Table \ref{uc} 
presents the information of the fuzzed programs and the number of unique crashes found in 24 hours.
From Table \ref{uc}, 
we have the following observations.
(1) 
For all the 13 programs,
V-Fuzz find the most unique crashes (the average number of unique crashes for one program is 1,114) 
and is much better than the other three fuzzers.
(2) 
Compared with VUzzer,
the average number of unique crashes found by V-Fuzz is improved by 35.8\%.
In addition,
for the program \texttt{cflow}, 
VUzzer did not find any crash while V-Fuzz found one crash.
(3) Compared with AFL,
there are five programs (\texttt{uniq}, \texttt{base64}, \texttt{who}, \texttt{pdf2svg} and \texttt{cflow}),
on which AFL has did not find any crash while V-Fuzz had a good performance.
(4) Compared with AFLFast,
there are also five programs (\texttt{uniq}, \texttt{base64},
\texttt{who}, 
\texttt{pdffonts} and \texttt{cflow}),
on which AFLFast did not find any crash while V-Fuzz did.

\begin{table}[!t]
\centering
\footnotesize
\caption{The number of unique crashes found for 24 hours.}
\setlength{\tabcolsep}{0.6mm}{
\begin{tabular}{llcccccccc}
\toprule[1.1pt]
\multirow{2}{*}{Application}&\multirow{2}{*}{Version}  &\multicolumn{4}{c}{Fuzzer}\\
\cline{3-6}
&& V-Fuzz & VUzzer & AFL & AFLFast \\
\midrule[1.1pt]
uniq & LAVA-M & 659 & 321 &0  &0 \\
base64 & LAVA-M & 128 & 100 &0 &0\\
who & LAVA-M & 117 & 92 & 0 &0 \\
pdftotext & xpdf-2.00 &  209 & 59 & 12 & 108\\
pdffonts & xpdf-2.00 & 581 & 367 & 13 & 0\\
pdftopbm & xpdf-2.00 & 50 & 25 & 37 & 35\\
\hline
\multirow{2}{*}{pdf2svg + libpoppler} & pdf2svg-0.2.3 & \multirow{2}{*}{3} & \multirow{2}{*}{2} &\multirow{2}{*}{0} &\multirow{2}{*}{1}\\
&libpoppler-0.24.5&\\
\hline
MP3Gain & 1.5.2 & 217 & 34 & 103 & 110\\
mpg321 & 0.3.2 & 321 &184 & 40 & 17 \\ 
xpstopng & libgxps-0.2.5 & 3,222 & 2,195 & 2 &2 \\
xpstops & libgxps-0.2.5 & 4,157 & 3,044 & 3 &3 \\
xpstojpeg & libgxps-0.2.5 & 4,828 & 4,243 &4 &4\\
cflow & 1.5 & 1& 0 & 0 & 0 &\\
\hline
\multicolumn{2}{c}{Total} &14,493  & 10,666  &214 & 280  \\
\hline
\multicolumn{2}{c}{Average} & 1,114 & 820 & 16 & 21 \\
\bottomrule[1.1pt]
\label{uc}
\end{tabular}}
\end{table}

\begin{figure*}[!t]
  \centering
  \includegraphics[width=4.3cm, height=2.8cm]{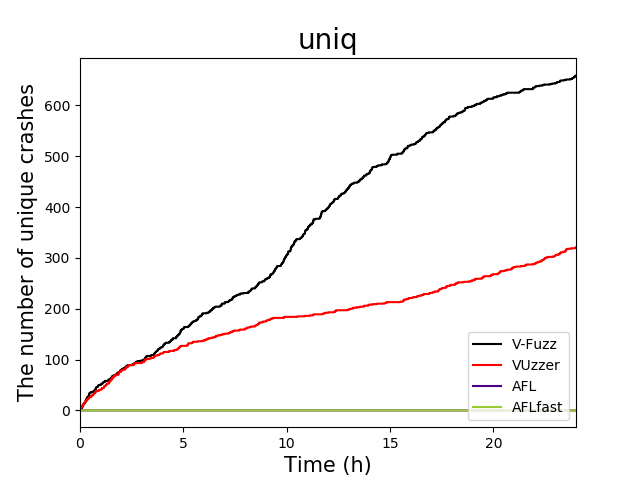}
  \includegraphics[width=4.3cm, height=2.8cm]{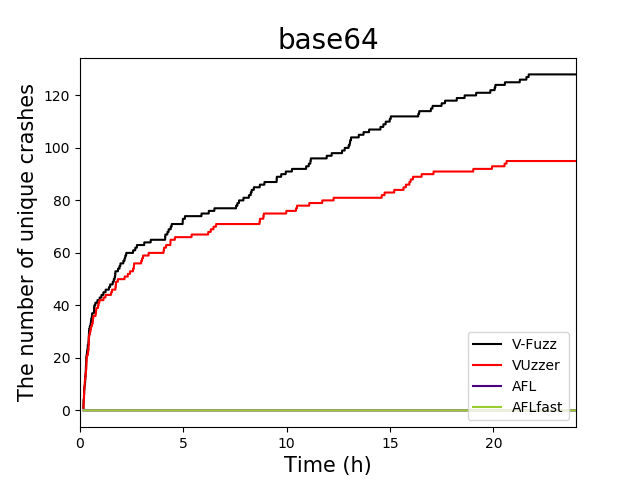}
  \includegraphics[width=4.3cm, height=2.8cm]{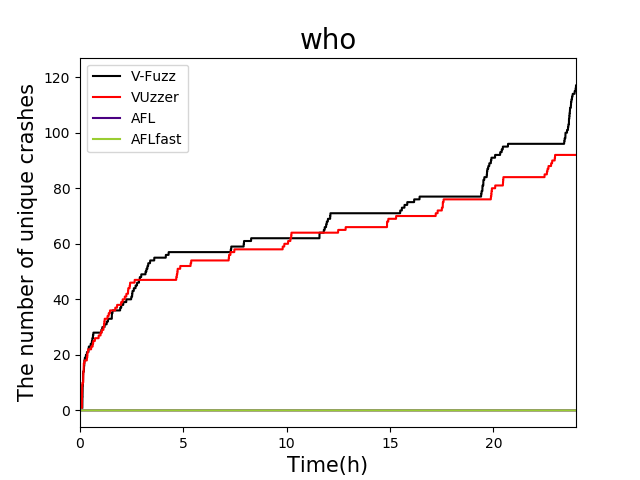}
  \includegraphics[width=4.3cm, height=2.8cm]{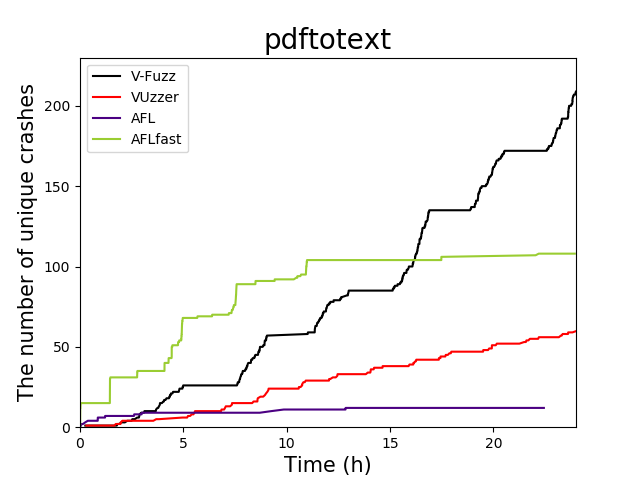}\\
  \includegraphics[width=4.3cm, height=2.8cm]{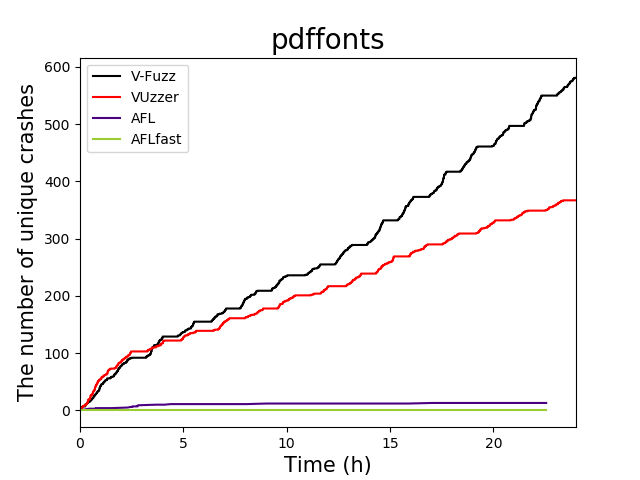}
   \includegraphics[width=4.3cm, height=2.8cm]{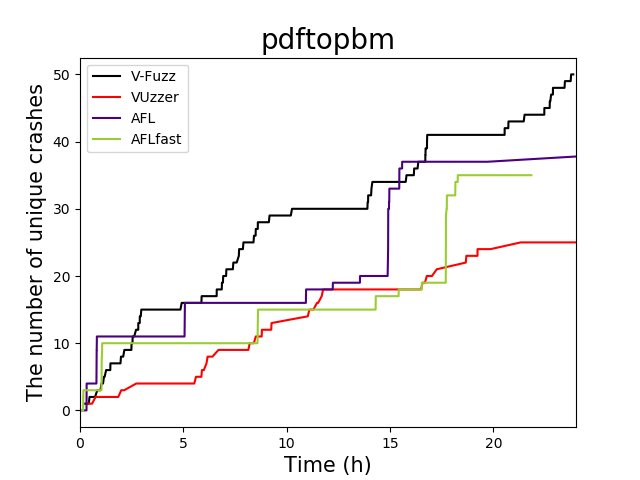}
   \includegraphics[width=4.3cm, height=2.8cm]{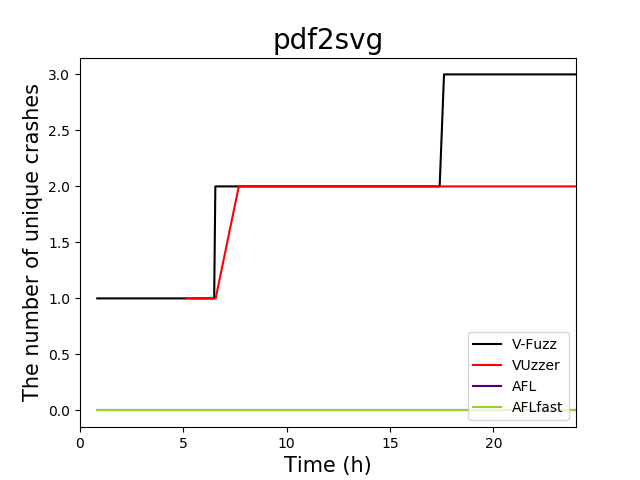}
  \includegraphics[width=4.3cm, height=2.8cm]{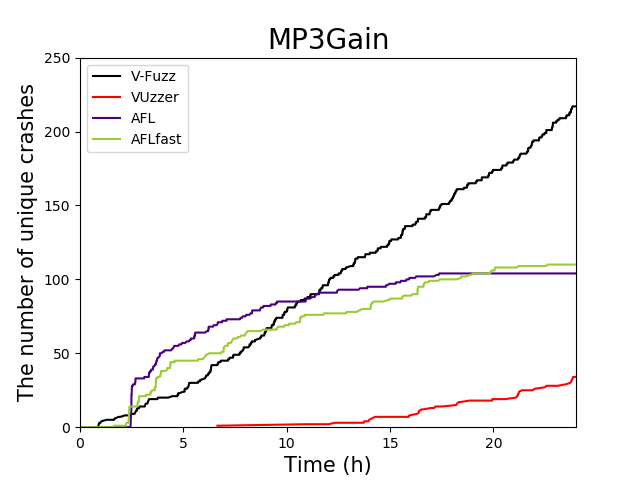}\\
  \includegraphics[width=3.5cm, height=2.8cm]{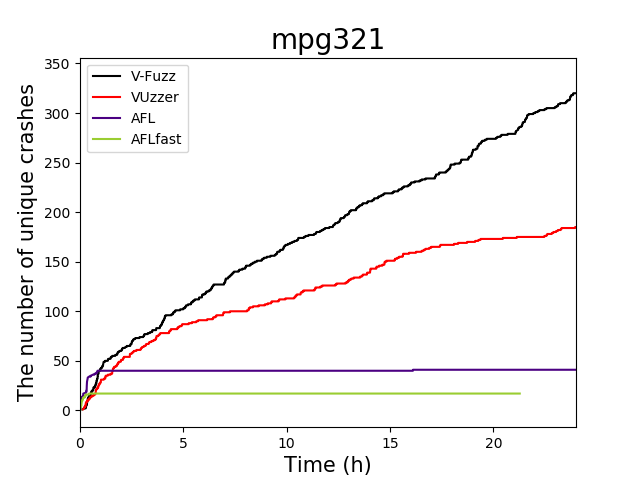}
  \includegraphics[width=3.5cm, height=2.8cm]{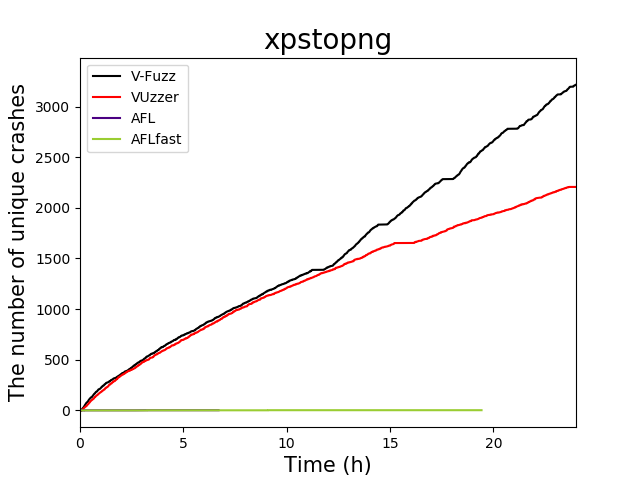}
  \includegraphics[width=3.5cm, height=2.8cm]{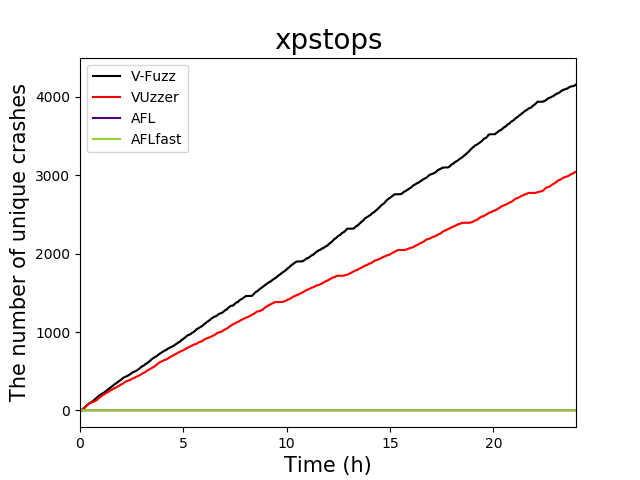}
  \includegraphics[width=3.5cm, height=2.8cm]{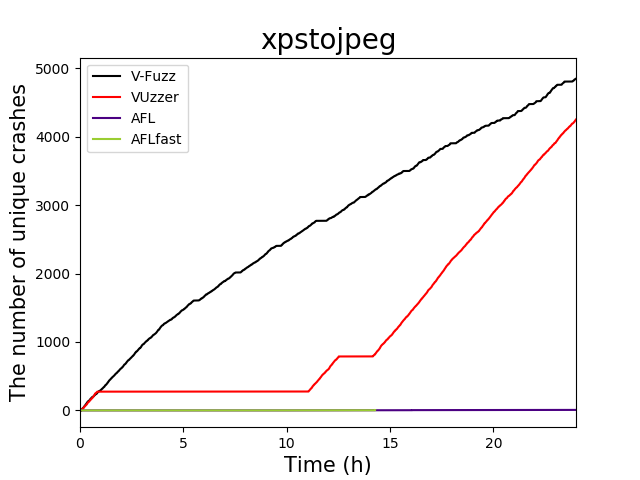}
  \includegraphics[width=3.5cm, height=2.8cm]{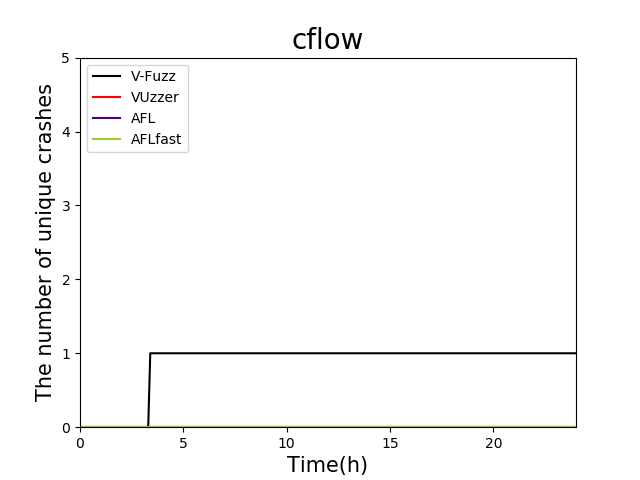}\\
  \caption{The growth curves of the number of unique crashes.}
  \label{f_uc}
\end{figure*}

\textbf{Whether V-Fuzz can find unique crashes quickly?}
Figure \ref{f_uc} shows the growth curves of the number of discovered unique crashes within 24 hours.
From the figure, 
we can observe that V-Fuzz finds unique crashes more quickly than other state-of-the-art fuzzers.
From the above results,
we can see that V-Fuzz has good performance in discovering unique crashes and outperforms state-of-the-art fuzzers.

\subsubsection{Vulnerability Discovery}

In this part,
we show V-Fuzz's capability of discovering vulnerabilities.
During the fuzzing test of V-Fuzz,
we collect the inputs which cause crashes.
For the three programs of LAVA-M,
we run the programs again with the crash inputs,
and verify the bugs they found.
Table ~\ref{LAVA-M-bug} shows the number of bugs found by V-Fuzz and VUzzer.
Each injected bug in LAVA-M has a unique ID,
and the corresponding ID is printed when the bug is triggered.
There are two kinds of bugs in LAVA-M:  listed and unlisted.
The listed bugs are those that the LAVA-M authors were able to trigger when creating the LAVA-M programs,
and the unlisted bugs are those that the LAVA-M authors were not able to trigger.
From Table ~\ref{LAVA-M-bug},
we can observe that V-Fuzz can trigger more bugs than VUzzer.
In addition, 
V-Fuzz is able to trigger several unlisted bugs and exhibits a better performance than VUzzer in this case too.
Table \ref{appendix-LAVA-M-id} shows the IDs of bugs triggered by V-Fuzz on three programs 
(\texttt{uniq}, \texttt{base64} and \texttt{who}) of LAVA-M datasets.

\begin{table}[!t]
\centering
\caption{The IDs of bugs triggered by V-Fuzz on LAVA-M.}
\setlength{\tabcolsep}{0.3mm}{
\begin{tabular}{lc}
\toprule[1pt]
Application & IDs of Bugs\\
\midrule[1pt]
\multirow{2}{*}{uniq} 
& 112, 130, 166, 169, 170, 171, 215, 222, 227, 293, 296\\
& 297, 321, 322, 346, 347, 368, 371, 372, 393, 396\\
& 397, 443, 446, 447, 468, 471, 472\\
\hline
\multirow{2}{*}{base64} 
& 1, 222, 235, 253, 255, 274, 276, 278, 284, 386, 521, 526\\
& 556, 558, 562, 576, 583, 584, 784, 790, 804, 805, 806, 813\\
& 832, 841, 843 \\
\hline
\multirow{5}{*}{who} 
& 1, 2, 3, 5, 6, 9, 14, 20, 22, 24, 56, 58, 60, 62, 75\\
& 77, 79, 81, 83, 87, 89, 109, 116, 124, 127, 129, 131\\
& 133, 137, 139, 143, 149, 151, 152, 153, 155, 157, 161\\
& 177, 179, 197, 1151, 1171, 1250, 1272, 1276, 1280, 1291\\
& 1721, 1783, 1880, 1888, 1908, 2486, 2947, 2957, 2979\\
& 3201, 3240, 3780, 3923, 3979 \\
\bottomrule[1pt]
\label{appendix-LAVA-M-id}
\end{tabular}}
\end{table}

For the real-world Linux applications,
in order to verify the vulnerabilities found by V-Fuzz,
we recompile the fuzzed programs with AddressSanitizer \cite{AS},
which is a memory error detector for C/C++ programs.
Then,
we execute these programs with the collected crash inputs.
AddressSanitizer can give the detailed information of the vulnerabilities.
Based on the information from AddressSanitizer,
we search the related information on the official CVE website  \cite{CVE}
and validate the vulnerabilities we found.

Table ~\ref{cve_found}
shows the detailed CVE information that we found.
We have found 10 CVEs in total,
which 3 of them
(CVE-2018-10767, CVE-2018-10733 and CVE-2018-10768) 
are newly found by us.
Moreover,
most of the fuzzed applications are shown to have CVEs.
The crash inputs which are found when fuzzing the programs of \texttt{xpdf-2.0} can also trigger the vulnerability of \texttt{xpdf-3.01}.
Finally,
most of the CVEs are buffer related errors,
which is reasonable as fuzzing test is good at finding this type of vulnerabilities.

\begin{table}[!t]
\centering
\small
\caption{The number of bugs found on LAVA-M.}
\setlength{\tabcolsep}{1.2mm}{
\begin{tabular}{ lccccccc}
\toprule[1.1pt]
\multirow{2}{*}{Application}&\multicolumn{3}{c}{V-Fuzz}&&\multicolumn{3}{c}{VUzzer} \\
\cline{2-8}
& Listed & Unlisted & Total & & Listed & Unlisted &Total\\
\midrule[1.1pt]
uniq &27 & 1  &28 && 26 &1 &27\\
base64 & 24 & 3 &27& & 23 &2 &25\\
who &57& 9 & 62 && 52 &7 &59\\
\bottomrule[1.1pt]
\label{LAVA-M-bug}
\end{tabular}}
\end{table}

\begin{table*}[!t]
\centering
\small
\caption{The CVEs found by V-Fuzz.}
\begin{tabular}{llllllll}
\toprule[1.1pt]
Application & Version & CVE & Vulnerability Typye\\
\midrule[1.1pt]
pdftotext & \multirow{3}{*}{xpdf$<=$3.01} & \multirow{3}{*}{CVE-2007-0104} & \multirow{3}{*}{Buffer errors}\\ 
pdffonts&&\\
pdftopbm&&\\
\hline
mpg321 & 0.3.2 & CVE-2017-11552 & Buffer errors \\
\hline
\multirow{5}{*}{MP3Gain} & \multirow{5}{*}{1.5.2} & CVE-2017-14406 & NULL pointer dereference\\
&  & CVE-2017-14407 & Stack-based buffer over-read\\
&  & CVE-2017-14409 & Buffer overflow\\ 
&  & CVE-2017-14410 & Buffer over-read &\\
&  & CVE-2017-12912 & Buffer errors\\
\hline
\multirow{2}{*}{libgxps} & \multirow{2}{*}{$<=$0.3.0} & CVE-2018-10767 (new) & Buffer errors\\
 &  & CVE-2018-10733 (new) & Stack-based buffer over-read\\
\hline
libpoppler & 0.24.5 & CVE-2018-10768 (new) & NULL pointer dereference\\
\bottomrule[1.1pt]
\label{cve_found}
\end{tabular}
\end{table*}

\subsubsection{Code Coverage}
V-Fuzz is neither a coverage-based fuzzer nor a directed fuzzer.
The goal of V-Fuzz is to find more bugs in a shorter period of time.
In this part, 
we show that V-Fuzz can achieve its goal without decreasing its code coverage.
The methods of calculating the code coverage for different fuzzers are different.
AFL and AFLFast utilize the static instrumentation with a compact bitmap to track edge coverage,
VUzzer leverages the dynamic binary instrumentation tool PIN 
 \cite{Luk2005Pin} 
to track basic block coverage.
Additionally, 
as  \cite{CollAFL} indicated,
most of the fuzzers cannot calculate code coverage accurately.
Therefore, 
for evaluating the code coverage,
we only compare V-Fuzz with VUzzer as they leverage the same approach to calculate the basic block coverage.

\begin{figure*}[!t]
  \centering
  \includegraphics[width=4.3cm, height=2.8cm]{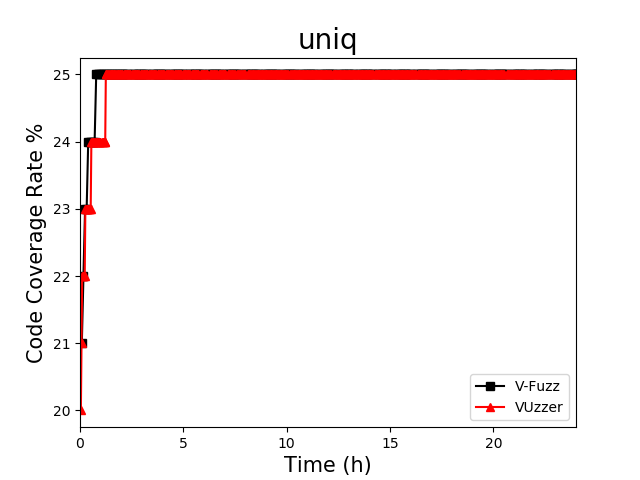}
  \includegraphics[width=4.3cm, height=2.8cm]{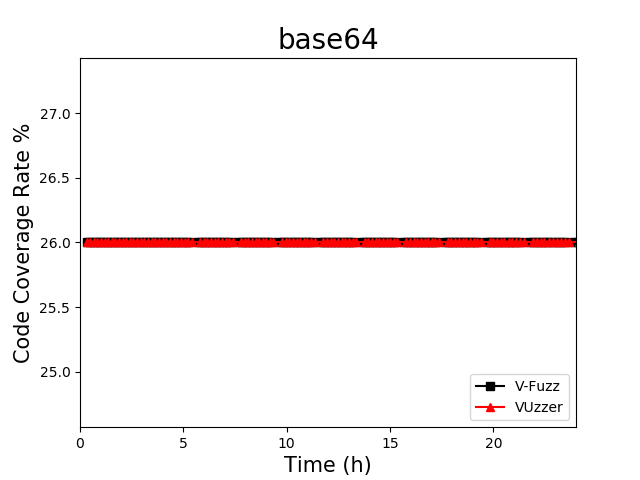}
  \includegraphics[width=4.3cm, height=2.8cm]{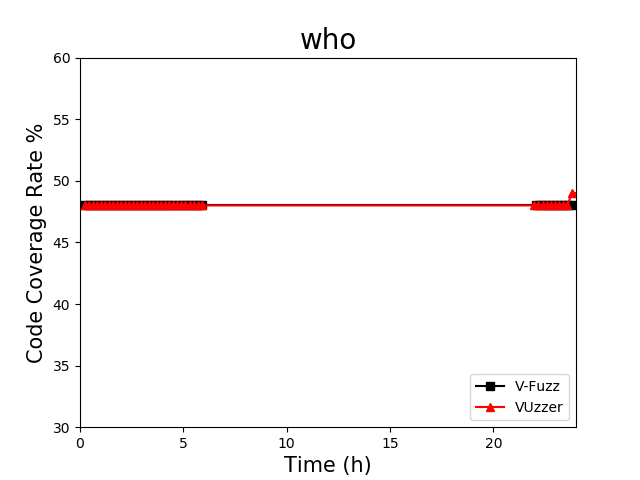}
  \includegraphics[width=4.3cm, height=2.8cm]{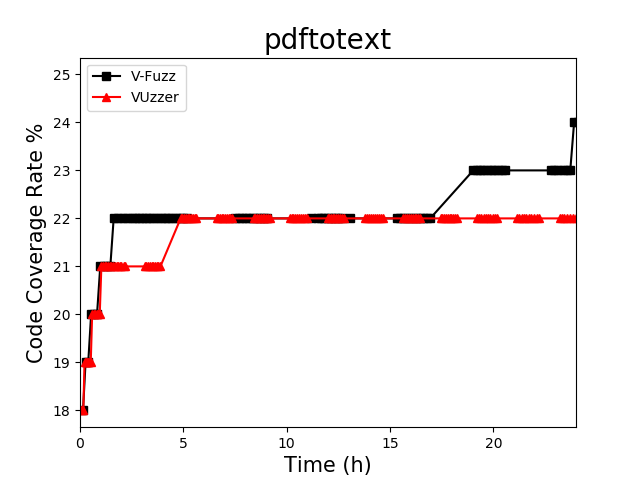}\\
  \includegraphics[width=4.3cm, height=2.8cm]{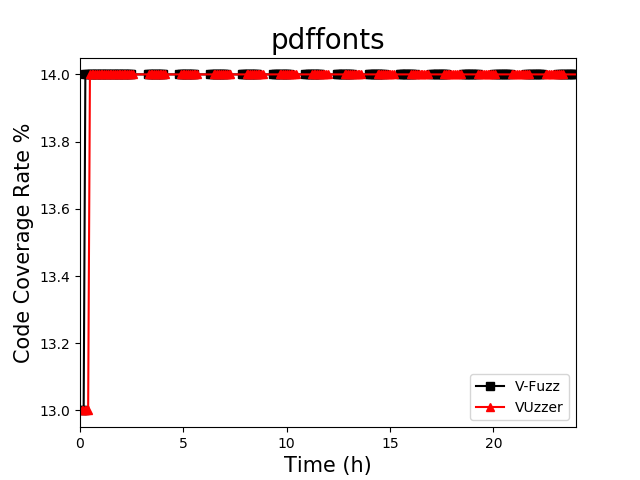}
   \includegraphics[width=4.3cm, height=2.8cm]{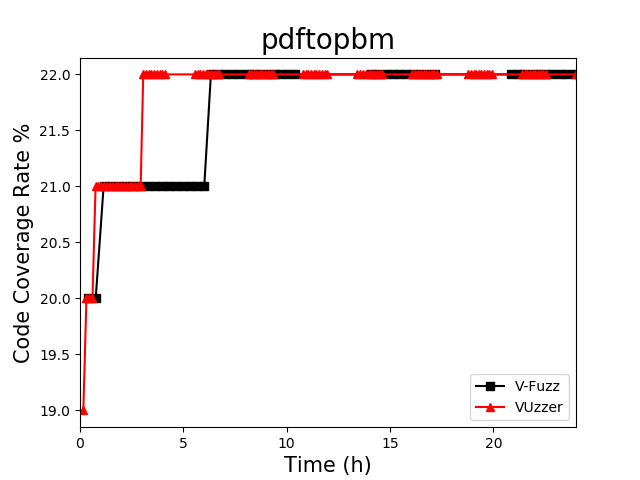}
   \includegraphics[width=4.3cm, height=2.8cm]{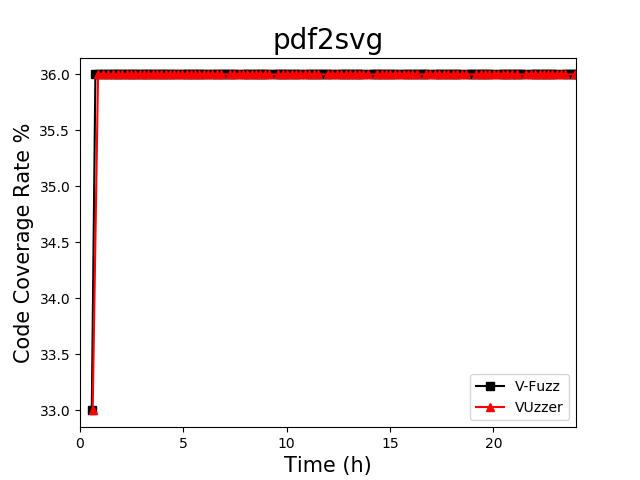}
  \includegraphics[width=4.3cm, height=2.8cm]{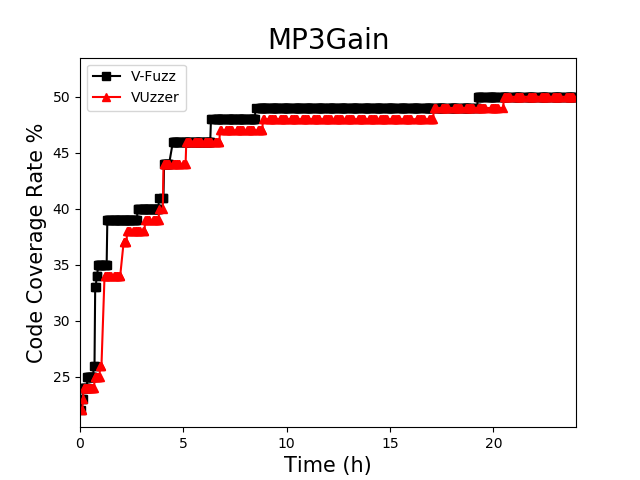}\\
  \includegraphics[width=3.5cm, height=2.8cm]{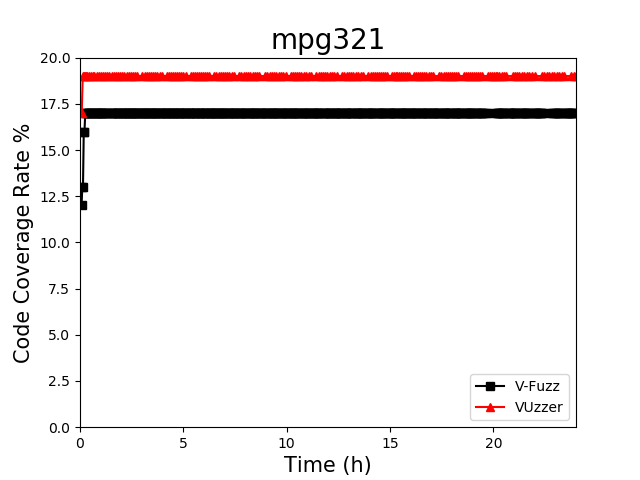}
  \includegraphics[width=3.5cm, height=2.8cm]{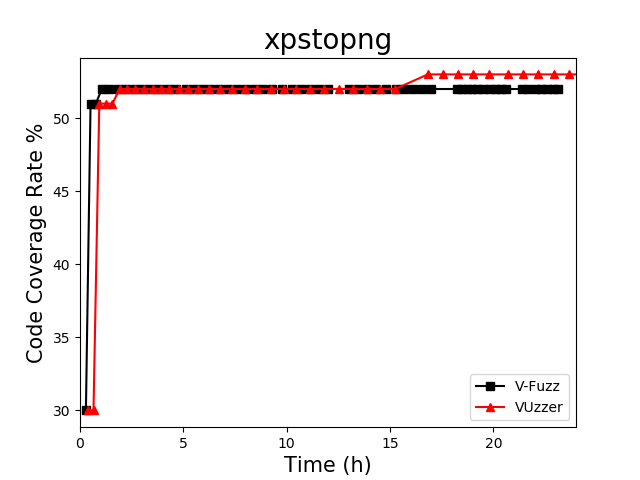}
  \includegraphics[width=3.5cm, height=2.8cm]{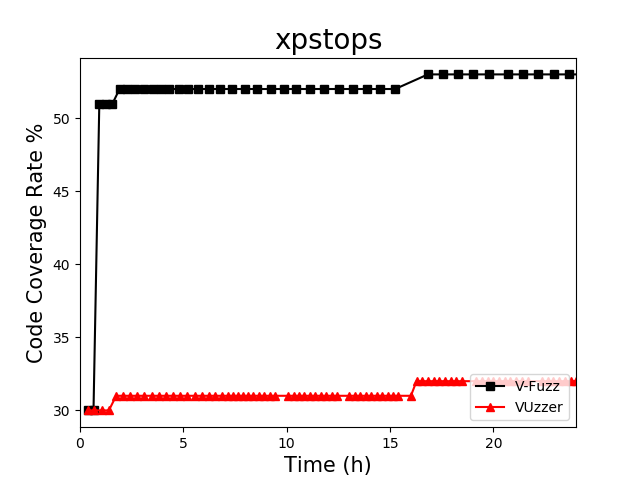}
  \includegraphics[width=3.5cm, height=2.8cm]{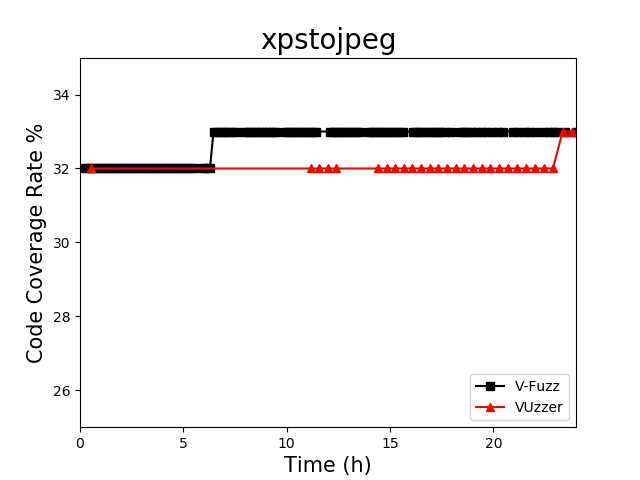}
  \includegraphics[width=3.5cm, height=2.8cm]{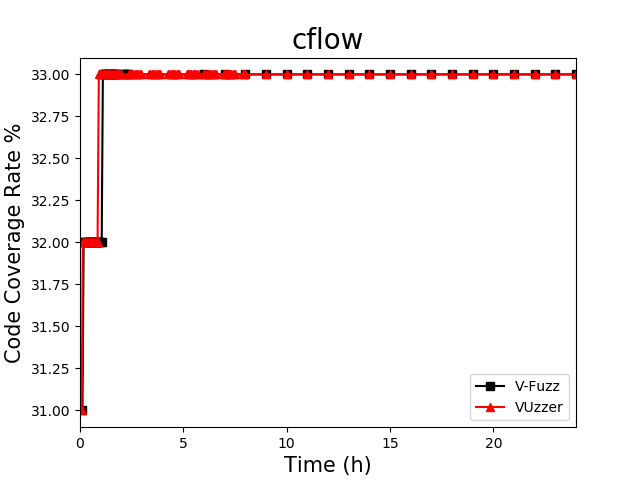}
  \caption{Code coverage rate.}
  \label{code_cov}
\end{figure*}

Figure \ref{code_cov} shows the growth curves of the code coverage within 24 hours for V-Fuzz and VUzzer.
From this figure,
we have the following observations:
(1) For most of the programs, 
the code coverage of V-Fuzz and VUzzer is almost the same.
(2) There is only one program (\texttt{mpg321}), 
on which the code coverage covered of VUzzer(19\%) is slightly larger than V-Fuzz(17\%).
(3) There are several programs (e.g., \texttt{MP3Gain}),
on which V-Fuzz has a higher code coverage,
especially for the program \texttt{xpstops}.

In conclusion,
although the code coverage is not the main objective that V-Fuzz focuses on,
the experiments show that V-Fuzz can achieve its main goal of finding crashes without decreasing much code coverage.
Moreover,
it can help reduce the false negative of the vulnerability prediction model.

\section{Discussion}

\subsection{Vulnerability-Oriented Fuzzing}

The blindness of fuzzing can be decreased by providing it with more information or combine fuzzing with other techniques.
For example, 
Driller  
\cite{Driller} is a popular fuzzer which combines fuzzing with symbolic execution techniques.
When the fuzzer got "stuck", 
the symbolic execution can compute a valid input for it.
However, 
as symbolic execution does not work well for real programs,
it is hard to apply this idea to real world.

The idea of vulnerability-oriented fuzzing is to improve a fuzzer's efficiency of finding vulnerabilities by assisting it with static vulnerability analysis techniques.
The static vulnerability analysis can provide a fuzzer with more information,
which can reduce the blindness of it.
It is worth noting that static vulnerability analysis can be conducted by traditional static vulnerability detection tools or AI techniques 
(e.g., deep learning or machine learning) 
based models.
Therefore, 
the vulnerability-oriented fuzzing can be implemented by many static analysis techniques.
In addition, 
the static analysis just gives an auxiliary information to a fuzzer,
which does not affect its extensibility.

Most of static analysis tools or models suffer from high false positive or false negative.
This problem has been plaguing the security researchers for many years, 
and there has been no good solution to solve it yet.
However,
static analysis can be combined with fuzzing to reduce its weakness.
Moreover,
unlike combining fuzzing with symbolic execution,
it has better extensibility and effectiveness.

\subsection{Limitations and Future Works}
Although V-Fuzz has improved the capability of fuzzing test in detecting vulnerabilities,
it still has several limitations.
First,
we only use the dataset from NIST 
\cite{NIST} to train and test the vulnerability prediction model.
The number of the labeled samples may not be sufficient,
which might further affect the performance of the model.
In the future,
we will pay more attention to collect and label more data to improve the model.
Second, 
our model is mainly designed for binary programs.
Although the model for predicting vulnerabilities for binary programs might be difficult than for source codes,
it is also necessary to design a model which can be applied to source codes directly.
In addition, 
as the instrumentation methods of most state-of-the-art fuzzers are based on compilers 
(e.g., gcc or clang), 
they need the source code of the fuzzed programs.
In order to combine our model with these fuzzers conveniently,
we will study source-code based prediction models,
and combine these models with more state-of-the-art fuzzers in the future.

\section{Related Work}

We introduce the related work from two perspectives: vulnerability prediction and fuzzing test.

\subsection{Vulnerability Prediction}

Here we introduce the related work about vulnerability prediction.
It needs to be emphasized that most of the previous work on vulnerability prediction is dealing with source code, 
which is different from us as we deal with binaries.


\subsubsection{Machine Learning based Vulnerability Prediction}

Machine learning based vulnerability prediction models are designed based on some software features which are manually selected or constructed by the experts.
Shin et al.
\cite{Shin2011Evaluating} 
leveraged complexity, code churn, and developer activity metrics to discriminate between vulnerable and neutral files. 
Hovsepyan et al.
\cite{Hovsepyan2012Software} presented a novel approach for vulnerability prediction that leverages on the analysis of raw source code as text.
Gegick et al.
\cite{Gegick2008Prioritizing} designed an attack-prone prediction model with some code-level metrics: static analysis tool alert density, code churn and count of the lines of code.
Neuhaus et al.
\cite{Neuhaus2007Predicting} proposed an approach to mine existing vulnerability databases and version archives automatically to map past vulnerabilities to components.
Shar et al.
\cite{Shar2013Mining} leveraged some static and dynamic attributes to predict the SQL injection and cross site scripting vulnerabilities.
Walden et al.
\cite{Walden2014Predicting} examined vulnerability prediction models for web applications,
and compared the performance of prediction models based on softwares metrics with that of models based on text mining.

\subsubsection{Deep Learning based Vulnerability Prediction}

Machine learning based approaches focus on manually designing features to represent a program.
However, these approaches still need too much energy.
Therefore, some deep learning based vulnerability prediction models are proposed. 
One of the biggest advantages of these models is the deep learning based models can learn features automatically according to different types of programs. 
Dam et al.
\cite{Dam2017Automatic} described a new approach built upon the Long Short Term Memory (LSTM) model to automatically learn both semantic and syntactic features in code.
Wang et al.
\cite{Wang2016Automatic} leveraged Deep Belief Network (DBN) to automatically learn semantic representation of programs from source code. 
Yang et al.
\cite{Yang2015Deep}
leveraged deep learning techniques to learn features for defect prediction.
Dam et al.
\cite{Dam2016DeepSoft} 
presented an end-to-end generic framework based on LSTM for modeling software and its development process to predict future risks. 
White et al.
\cite{White2015Toward} demonstrated that the deep software language models have better performance than n-grams based models. 
Gu et al.
\cite{Gu2016Deep} proposed a deep learning based approach to generate API usage sequences for a given natural language query.
Huo et al.
\cite{Huo2016Learning} proposed a convolutional neural network based model to learn unified features from natural language and source code in programs for locating the potential buggy source code.

\subsection{Fuzzing Test}

\subsubsection{Symbolic Execution based Fuzzing.}
Most of whitebox fuzzers leverage symbolic execution or
concolic execution
(combines symbolic execution and concrete execution) 
to generate inputs that can execute new paths.
SAGE  
\cite{Godefroid2008Automated}
is a whitebox fuzzer,
which uses symbolic execution to gather path constraints of conditional statements.
CUTE  
\cite{sen2005cute}
is a unit testing engine for C programs by combining symbolic and concrete execution.
ZESTI  
\cite{marinescu2012make}
is a software testing tool that takes a lightweight symbolic execution mechanism to execute regression test.
Dowser 
\cite{Haller2013Dowsing}
is a guided fuzzer that combines taint tracking, 
program analysis and symbolic execution to detect buffer overflow vulnerabilities.
Driller  
\cite{Driller} 
combines AFL with concolic execution to generate inputs that can trigger deeper bugs.
SmartFuzz 
\cite{Molnar2009Dynamic} 
focuses on discovering integer bugs for x86 binaries using symbolic execution.

\subsubsection{Taint Analysis based Fuzzing.}
Another common technique that is leveraged by fuzzing is taint analysis,
especially dynamic taint analysis.
BuzzFuzz 
\cite{Ganesh2009Taint}
is an automated whitebox fuzzer.
It employs dynamic taint tracing to locate the regions of the original inputs that 
influence values used at key attack points.
TaintScope 
\cite{wang2010taintscope}
is a fuzzing system that uses dynamic taint analysis and symbolic execution to fuzz x86 binary programs.

\subsubsection{Heuristic Algorithm based Fuzzing.}
Most mutation-based fuzzers employ heuristic algorithms such as evolutionary algorithms to guide them select and generate high-quality inputs.
AFL  
\cite{AFL} is the most popular mutation-based fuzzer which leverages a simple evolutionary algorithm to generate inputs.
AFLFast  
\cite{Pham2016Coverage}
is a coverage-based greybox fuzzer.
It leverages a Markov chain model to generate inputs that tends to arrive at the ``low-frequency" paths.
VUzzer  
\cite{Rawat2017VUzzer}
is an application-aware fuzzer.
It leverages an evolutionary algorithm to generate inputs that can discover bugs in deep paths.
AFLGo 
\cite{AFLGo}
is a directed greybox fuzzer,
which generates inputs with the aim of reaching some specific locations by a simulated annealing-based power schedule approach.
SlowFuzz 
\cite{Petsios2017SlowFuzz}
is a domain-independent framework for automatically finding algorithmic complexity vulnerabilities.

\subsubsection{Machine Learning based Fuzzing.}
With the development of artificial intelligence techniques,
there are some researches that begin to explore how to apply these techniques into fuzzing test.
Godefroid et al.  
\cite{godefroid2017learn}
propose a novel fuzzing method, 
which uses a learnt input probability distribution to intelligently guide where to fuzz inputs.
Rajpal et al.  
\cite{rajpal2017not}
present a learning approach that uses neural networks to learn patterns in the inputs files to guide future fuzzing.
Nichols et al.  
\cite{nichols2017faster}
propose a method that uses Generative Adversarial Network (GAN) models to reinitialize the seed files to improve the performance of fuzzing.
This paper 
\cite{bottinger2018deep}
formalizes fuzzing as a reinforcement learning problem by using the concept of Markov decision process.

\subsubsection{Other Fuzzing Researches.}
kAFL 
\cite{kAFL} is a hardware-assisted feedback fuzzer that focuses on fuzzing x86-64 kernels.
IMF 
\cite{han2017imf} leverages inferred dependence model to fuzz commodity OS kernels.
Skyfire 
\cite{wang2017skyfire} leverages the knowledge of existing samples to generate well-distributed inputs for fuzzing.
CollAFL  
\cite{CollAFL} is a coverage sensitive fuzzing solution that mitigates path collisions by improving the accurate coverage information.
T-Fuzz  
\cite{T-Fuzz} leverages a lightweight dynamic tracing-based approach to infer all checks that could not be passed and generates mutated programs where the checks are negated.
Angora  
\cite{Chen2018Angora} is an mutation-based fuzzer that increases branch coverage by solving path constraints without symbolic execution.

\section{Conclusion}
In this paper,
we design and implement V-Fuzz,
a vulnerability-oriented evolutionary fuzzing framework.
By combining the vulnerability prediction with evolutionary fuzzing,
V-Fuzz can generate inputs that tend to arrive at the potential vulnerable regions.
We evaluate V-Fuzz on popular benchmark programs (e.g., \texttt{uniq}) of LAVA-M  
\cite{Dolangavitt2016LAVA},
and a variety of real-world Linux applications including the audio processing softwares (e.g., \texttt{MP3Gain}), 
pdf transformation tools (e.g., \texttt{pdftotext}) and xps documents library (e.g., libgxps).
Compared with the state-of-the-art fuzzers,
the experimental results demonstrate that V-Fuzz can find more vulnerabilities quickly.
In addition,
V-Fuzz has discovered 10 CVEs,
and 3 of them are newly discovered.
We reported the new CVEs, 
and they have been confirmed and fixed.
In the future,
we will study to leverage more advanced program analysis techniques to assist fuzzer in discovering vulnerabilities.

\end{document}